\documentclass[preprint]{aastex}
\usepackage{epsf}
\usepackage{graphicx}

\newcommand{\bfm}[1]{\mathit{\mathbf{#1}}}

\shorttitle{Unified Treatment of Open and Closed Topologies}
\shortauthors{Oran et al.}

\begin{document}

\title{A Global Wave-Driven MHD Solar Model with a Unified Treatment of Open and Closed Magnetic Field Topologies}

\author{R. Oran\altaffilmark{1}, B. van der Holst\altaffilmark{1}, E. Landi\altaffilmark{1}, M. Jin\altaffilmark{1},  I. V. Sokolov\altaffilmark{1} and  T. I. Gombosi\altaffilmark{1}}

\email{oran@umich.edu}

\altaffiltext{1}{Atmospheric, Oceanic and Atmospheric Sciences, University of Michigan, 2455 Hayward, Ann Arbor, MI, 48105, USA}

\begin{abstract}
We describe, analyze and validate the recently developed Alfv{\'e}n Wave Solar Model (AWSoM), a 3D global model starting from the top of the chromosphere and extending into interplanetary space (up to 1-2 AU). This model solves the extended two-temperature magnetohydrodynamics equations coupled to a wave kinetic equation for low frequency Alfv\'en waves. In this picture, heating and acceleration of the plasma are due to wave dissipation and wave pressure gradients, respectively. The dissipation process is described by a fully developed turbulent cascade of counter-propagating waves. We adopt a unified approach for calculating the wave dissipation in both open and closed magnetic field lines, allowing for a self-consistent treatment of any magnetic topology. 
Wave dissipation is the only heating mechanism assumed in the model, and no geometric heating functions are invoked. Electron heat conduction and radiative cooling are also included. We demonstrate that the large-scale, steady-state (in the co-rotating frame) properties of the solar environment are reproduced, using three adjustable parameters: the Poynting flux of chromospheric Alfv{\'e}n waves, the perpendicular correlation length of the turbulence, and a pseudo-reflection coefficient. We compare model results for Carrington Rotation 2063 (November-December 2007) to remote observations in the EUV and X-ray ranges from STEREO, SOHO and Hinode spacecraft, as well as to in-situ measurements performed by Ulysses. The model results are in good agreement with observations. This is the first global model capable of simultaneously reproducing the multi-wavelength observations of the lower corona and the wind structure beyond Earth's orbit.
\end{abstract}

\keywords{magnetohydrodynamics (MHD) - Sun:corona - Sun:solar wind - method:numerical - turbulence}

\section{Introduction}
The solar corona is the extension of the Sun's atmosphere, made of ionized plasma that is organized by the solar magnetic field into distinct structures. Away from the Sun, coronal plasma is accelerated radially outward, eventually forming the solar wind, which extends into interplanetary space. The corona-wind system is highly dynamic. The overall topology slowly changes throughout the 11-year long solar cycle. On shorter time scales, localized eruptions in the solar atmosphere such as flares and coronal mass ejections (CME's) may occur, releasing energy into the system and interacting with the ambient plasma environment. Understanding the processes dominating the ambient solar corona and solar wind is a crucial step in the study and prediction of space weather.\par In the last few decades, considerable effort was made to better understand this system, with an ever increasing availability of observations, as well as the development of complex computational models. This allowed for significant progress to be made, but a number of questions still remain unanswered, mainly relating to the global distribution of coronal temperatures and solar wind speeds. Coronal plasma can reach temperatures of 1-2MK, which is about two orders of magnitude higher than the temperature in the underlying chromosphere. Several promising theories were put forth in order to explain this sharp gradient, invoking either wave turbulence or magnetic loop reconnection as the source of energy (c.f. \cite{Karachik2011}). The corona exhibits a global variability of temperature and density between the different magnetic structures (e.g. open field lines, quiet-Sun closed loops, and active regions). The challenge of explaining the observed distribution of heating rates in this complex magnetic topology is still not fully addressed. In addition, the magnetic field topology can be related to the structure of the solar wind. The wind can be classified into two distinct flow types, differing in wind speed, temperature and heavy ion composition. Commonly known as the fast and slow solar wind, these flow types have been persistently detected by spacecraft at Earth's orbit and in polar orbits around the Sun \citep{McComas2000,mccomas2007,Ebert2009}. Although there is still an on-going debate regarding the source regions and acceleration mechanisms of the slow solar wind \citep{kohl2006,suess2009,abbo2010,antiochos2011,antonucci2011}, it is commonly agreed that the high-latitude fast solar wind emanates from regions of open magnetic field lines. A solar model with a realistic magnetic field should be able to at least reproduce the ambient large-scale distribution of flow speeds.\par 
Current state of the art models of the solar environment can be divided into two major types: 1. Ideal models which prescribe the magnetic field topology (usually that of an ideal polar coronal hole), allowing for a detailed description of the physical mechanisms  involved in coronal heating or wind acceleration. 2. Realistic global (3D) models which use the measured photospheric magnetic field as input. The latter use a more simplified approach, often invoking empirically-motivated heating functions. Although models of the first type can be used to gain a deeper physical insight into the dynamics of specific ideal structures, global models allow us to test our theories against time-dependent observations of the highly variable solar environment. Furthermore, the study of space weather prediction relies on the development of robust and validated global models, which are capable of reproducing as large a range of observables as possible. These include the density and temperature distributions in the solar corona, as well as the flow properties of the solar wind. As these are organized by the magnetic field, global models should be able to treat any arbitrary magnetic topology and to allow open and closed field line structures to develop self-consistently with the plasma. In this work we present, analyze and validate the results from the Alfv{\'e}n Wave Solar Model (AWSoM), a global MHD model driven by Alfv{\'e}nic turbulence, extending from the upper chromosphere into the solar wind. The model applies a unified treatment of wave dissipation in open and closed field lines, as we described in our earlier paper \citep{sokolov2013}. In this work we analyze our choice of wave dissipation and its implications, extend the model to the solar wind beyond Earth's orbit, and validate our results against remote and in-situ observations. We discuss the current state of wave-driven global modeling and the need for the present model below. \par 
\subsection{The Challenge of Global Solar Modeling}\label{S:global}
Early 3D models of the corona were based on a potential field extrapolation of the measured photospheric magnetic field (e.g. \citet{altschuler1969}). Although these models were successful in predicting the location of major topological features of the corona, such as helmet streamers and coronal holes, their assumptions were restrictive and known to be physically invalid in the corona, where a variety of current systems occur.\par
First attempts at a self-consistent 3D model were based on ideal magnetohydrodynamics (MHD). There are a number of numerical models that solve the MHD equations using the photospheric magnetic field as an inner boundary condition \citep{usmanov1993,linker1999,mikic1999,roussev2003,Riley2006,cohen2007}. These models invoked empirical source terms in the MHD equations (e.g. geometric heating/cooling functions, variable polytropic index) in order to reproduce the observed properties of the solar corona and solar wind. Although they capture the overall solar wind structure, they give less accurate results in the presence of CME's and shocks. In these models the inner boundary is set at the bottom of the corona with temperatures in the 1MK range, thus avoiding the problem of the formation of the hot corona from the much cooler chromosphere.\par
Other models have addressed coronal heating by setting the lower boundary at the top of the chromosphere and invoking more detailed heating functions \citep{lionello2009,downs2010}. These models were the first to include the transition region in a global 3D model and produce simulated EUV images of the lower corona. By comparing the simulated EUV emission to observations one could test the temperature and density distributions predicted by the model. Although successful in reproducing key features in the observations, they were restricted by the use of heating functions that were not self-consistently coupled to the plasma. Furthermore, different geometric heating functions had to be introduced in order to account for the different heating rates observed in coronal holes, streamer belts and active regions. It should also be noted that these efforts were focused on reproducing the observed emission from the lower corona, and did not attempt to predict the distribution of solar wind properties in the heliosphere.\par
\subsection{Wave-Driven Solar Models}
Turbulent Alfv{\'e}n waves emanating from the chromosphere were suggested as a possible mechanism to both accelerate the solar wind and heat the plasma in the solar corona \citep{alazraki1971,belcher1971}. The source of such wave energy is assumed to be the constant reconfiguration of the magnetic field in the photosphere and chromosphere. In this picture, the plasma is accelerated due to gradients in the wave pressure, while heating is achieved by dissipation of wave energy. The importance of wave-driven models was further demonstrated in \citet{evans2008}, who compared the Alfv{\'e}n speed profiles predicted by ten different solar models against type II radio bursts observations. The survey included 1D wave-driven models, global MHD models with ad-hoc heating functions, as well as empirically derived profiles. The authors found that global MHD models with ad-hoc heating terms were less consistent with observations in the lower corona compared to wave-driven models.\par 
The key element of any wave driven model is the exchange of momentum and energy between the plasma and the Alfv{\'e}n wave field. This interaction can be described self-consistently by the coupled system of the MHD and a wave kinetic equation for Alfv{\'e}n waves. The latter can be derived under the Wentzel-Kramers-Brillouin (WKB) approximation \citep{jacques1977}, which is valid for wave-lengths shorter than the length scale of variations in the background. This description is justified if one wishes to describe the large scale dynamics of the system, rather than the detailed conversion of wave energy into thermal energy.\par  
The coupled system of MHD equations and a wave kinetic equation were first solved for an axisymmetric solar wind model in \citet{Usmanov2000}, and later in \citet{Usmanov2003}, assuming an ideal dipole magnetic field. The model results were in general good agreement with Ulysses observations of the fast and slow solar wind. However, this model did not address the problem of coronal heating, since the inner boundary was already at the 1MK range. In addition, the description only accounted for Alfv{\'e}n waves of a single polarity, and their dissipation was described by a simple linear loss term.\par 
Inhomogeneities in the magnetic field and plasma parameters can cause Alfv\'en waves to undergo reflections, giving rise to counter-propagating waves. Counter-propagating waves will also naturally occur independently of reflections along closed-field lines, where outgoing waves of opposite polarities are launched from the two foot points. Regardless of their formation mechanism, counter-propagating waves will undergo non-linear wave-wave interactions and subsequent evolution of the wave spectrum. In a turbulent regime, this scenario will lead to an energy cascade into smaller and smaller wave lengths, a process that must eventually result in the conversion of wave energy into plasma thermal energy. Previous works have simulated this process directly by describing wave reflections and frequency-dependent wave-wave interactions in idealized open flux tubes (e.g. \citet{cranmer2005}, \citet{cranmer2007}, \citet{verdini2007}). While this approach is instructive for prescribed magnetic fields, its application to a 3D model with a realistic and self-consistent magnetic field is quite involved.\par 
An alternative to this approach was proposed in \citet{hollweg1986}, who calculated a Kolmogorov-type energy dissipation rate. In this approach, the cascade process due to the presence of counter-propagating waves was assumed to be fast enough, such that the reflected wave energy is totally dissipated before it can propagate away. Under this assumption one can relate the dissipation rate to the macroscopic properties of the system. This property of the Kolmogorov-type treatment makes it especially attractive for global MHD modeling, since it does not require directly describing wave reflections and spectral evolution. However, it does require us to make some assumptions about the efficiency of the cascade process, as we will discuss in Section \ref{S:dissip}.\par 
The dissipation rate proposed in \citet{hollweg1986} was applied to a magnetogram-driven coronal model in \citet{vanderholst2010}, which also included separate electron and proton temperatures. The model results were validated against observations at 1AU in \citet{jin2012}. In \citet{evans2012}, this model was extended to include the contribution of surface Alfv{\'e}n waves to the dissipated energy. However, Hollweg's approach was developed for the case where outgoing waves of a single polarity are injected into the base an expending flux tube. Thus it could not be applied to closed field lines, and consequently no wave energy was injected at the foot points of coronal loops. It should be noted that the \citet{vanderholst2010,evans2012} model did not aim to create the global structure of the corona starting from the rather uniform underlying chromosphere. Instead, it derived the temperature and density distribution at the inner boundary from tomographic data of the ~1MK coronal plasma. 
\subsection{The AWSoM Model Approach} 
In this work, we describe the Alfv{\'e}n Wave driven Solar Model (AWSoM), a first-principles global model extending from the top of the chromosphere out to the solar wind, based on a wave kinetic / extended MHD framework. The model is driven by a Poynting flux of Alfv{\'e}n waves that is injected at all magnetic field foot points, and its magnitude is related to the local radial magnetic field. The wave energy is then transported along magnetic field lines into the corona and the solar wind.\par 
In order to create the observed temperature and density distributions without invoking geometric heating function, we require a heating mechanism that depends on the magnetic field topology. At the same time, the open and closed field line regions should emerge automatically, without the need to a-priori determine their locations. In our earlier paper of \citet{sokolov2013}, the Kolmogorov-type approach presented in \citet{hollweg1986} for calculating the wave dissipation in open flux tubes was generalized such that it can also be applied to closed magnetic field lines, where counter propagating waves naturally arise from the topology. \par 
In order to complete the description of a wave driven model, one must also specify the Poynting flux injected into the system. \citet{suzuki2006} showed that the required flux can be determined by considering energy conservation along expanding flux tubes in the solar wind, i.e. by relating the energy flux at the foot-point of a field line to the final wind speed along the same field line. One clear limitation of this approach is that it can only be applied to open field lines. Furthermore, complete information about the final wind speed even for all open field lines is not available, and the terminal wind speed at a spherical surface at 1AU has to be taken from some semi-empirical model, e.g. the Wang-Sheeley-Arge (WSA) model \citep{arge2000}. In this work, we wish to take a different approach, and specify a wave Poynting flux that is independent of conditions at 1AU and only constrained by observations of chromospheric Alfv{\'e}n waves.\par
This paper is organized as follows. The AWSoM model equations and the physical processes included in the model, as well as the constraints on the adjustable input parameters are described in Section \ref{S:Description}. The numerical model is described in Section \ref{S:numerical}, where we discuss the choice of computational grid and the inner boundary conditions. We then present results from idealized simulations in Section \ref{S:ideal}, where we focus on analyzing the validity and implications of our choice of wave dissipation. Model validation for a real magnetogram field for a solar minimum case is presented in Section \ref{S:minimum}. We compare our model prediction to remote observations of the solar corona (line-of-sight EUV and X ray images) and in situ observations made by the Ulysses spacecraft. This enables us to test how well the model reproduced both coronal structures and the fast and slow solar wind distribution. Conclusions and discussion of the results and future work can be found in Section \ref{S:discussion}.
\section{Model Description}\label{S:Description}
\subsection{Governing Equations}\label{S:equations}
The macroscopic evolution of the coronal and the solar wind plasma can be adequately described by the equations of non-resistive MHD. Although this approximation breaks down in the partially ionized chromosphere, by setting the inner boundary of the model at the top of the chromosphere, resistive effects can be neglected. To account for the different thermodynamic processes acting on electrons and protons, we start from the two-temperature MHD equations derived in \citet{braginskii1965}. We assume that the Hall effect can be neglected, and that the electrons and protons flow with the same velocity. Together with the assumption of quasi-neutrality this leads to single-fluid continuity and momentum equations. The electrons and protons obey separate energy equations. Non ideal-MHD processes such as heating, electron heat conduction and radiative cooling become important at certain regions and should be included as source terms in the energy equations. Finally, the modified MHD equations are coupled to wave kinetic equations for parallel and anti-parallel waves, as described in \citet{sokolov2009}, \citet{vanderholst2010} and \citet{sokolov2013}. The governing equations then become: 
\begin{equation}\label{eq:mass}
\frac{\partial \rho}{\partial t} +\bfm{\nabla \cdot} (\rho \bfm{u})=0,
\end{equation}

\begin{equation}\label{eq:mom}
\rho\frac{\partial \bfm{u}}{\partial t}+\rho\bfm{u\cdot\nabla}\bfm{u}=-\rho\frac{GM_\odot}{r^3}\bfm{r}-\nabla(p_e+p_p+p_w)+\frac{1}{\mu _0}(\bfm{\nabla\times B})\bfm{\times B},
\end{equation}

\begin{equation}\label{eq:induction}
\frac{\partial \bfm{B}}{\partial t}+\bfm{\nabla\cdot}(\bfm{u}\bfm{B}-\bfm{B}\bfm{u})=0 ,
\end{equation}

\begin{equation}\label{eq:wavekinetic}
\frac{\partial w^\pm}{\partial t}+\bfm{\nabla\cdot} [(\bfm{u}\pm\bfm{V_A})w^\pm]=-\frac{1}{2}(\bfm{\nabla\cdot}\bfm{u})w^\pm - Q_w^\pm,
\end{equation}

\begin{equation}\label{eq:protons}
\frac{\partial p_p}{\partial t}+\bfm{\nabla\cdot}(p_p\bfm{u})=(\gamma-1)[-p_p\bfm{\nabla\cdot}\bfm{u}+ \frac{1}{\tau_{pe}}(p_e-p_p)+f_p Q_w],
\end{equation}

\begin{equation}\label{eq:electrons}
\frac{\partial p_e}{\partial t}+\bfm{\nabla\cdot}(p_e\bfm{u})=(\gamma-1)[-p_e\bfm{\nabla\cdot u}+\frac{1}{\tau_{pe}}(p_p-p_e) - Q_{rad}+(1-f_p)Q_w -\bfm{\nabla\cdot q}_e].
\end{equation}

The basic state variables are the mass density, $\rho$, the bulk flow velocity, $\bfm{u}$, the magnetic field, $\bfm{B}$, and the proton and electron thermal pressures, $p_p$ and $p_e$, respectively. $w^\pm$ is the energy density of Alfv\'en waves propagating parallel(+) or anti parallel(-) to the magnetic field. Next, $G$ is the gravitational constant, $M_\odot$ is the solar mass, $\mu_0$ is the magnetic permeability, and $\gamma$ the polytropic index set to be constant at $5/3$.  The Alfv\'en velocity is given by $\bfm{V}_A=\bfm{B}/\sqrt{\mu_0 \rho}$. For the wave pressure tensor, we use the derivation by \citet{jacques1977}, who found it to be isotropic and given by $p_w = (w^+ +w^ -)/2$.\par 
Eqs. (\ref{eq:mass})-(\ref{eq:mom}) describe the conservation of mass and momentum. Eq. (\ref{eq:mom}) includes acceleration due to solar gravity, gradients in the electron, proton and wave pressures and the Lorentz force. Eq. (\ref{eq:induction}) is the induction equation for the magnetic field in the non-resistive limit. The wave kinetic equations are given in Eq. (\ref{eq:wavekinetic}), which represents two separate equations, for waves traveling parallel and anti parallel to the magnetic field. The wave energy density dissipation rate for each wave polarity is denoted by $Q_w^\pm$. The total wave energy density dissipation rate is given by $Q_w = Q_w^+ + Q_w^-$. The explicit form of the dissipation term will be discussed in Section \ref{S:dissip}.\par
The pressure equations for protons and electrons are given in Eqs. (\ref{eq:protons}) and (\ref{eq:electrons}). Both equations include electrons-protons heat exchange and the total wave dissipation rate, $Q_w$. The radiative cooling rate, $Q_{rad}$, is assumed to be due to electronic de-excitation, which becomes important in the cooler lower corona. The cooling rates are calculated from the CHIANTI 7.1 atomic database \citep{dere1997,landi2013}, where the ion population is determined by assuming coronal elemental abundances (taken from \citet{feldman1992}) and ionization equilibrium (obtained from the ionization and recombination rates appearing in \citet{landi2013}).\par
The total dissipated wave energy heats both protons and electrons, with the fraction of heating going into the protons denoted by the constant $f_p=0.6$  (see \citet{breech2009}, \citet{cranmer2009} for more details). Heat exchange due to Coulomb collisions between electrons and protons enters the energy equations through the second term on the right hand side of both equations. The collisional heat exchange results in temperature equilibration on a time scale $\tau_{pe}$, which is given by \citep{goedbloed2004}:
\begin{equation}
\tau_{pe}=3\pi \sqrt{2\pi}\epsilon_0\frac{m_p}{\sqrt{m_e}}\frac{(kT_e)^{3/2}}{ln\Lambda e^4 n},
\end{equation}
where $m_p$ and $m_e$ are the proton and electron masses, respectively, $e$ is the elementary charge, $k$ is the Boltzmann constant, $T_e$ is the electron temperature, $\epsilon_0$ is the permittivity of free space, $n$ is the plasma number density (under the assumption of quasi-neutrality) and $\ln\Lambda$ is the Coulomb logarithm, taken to be uniform with $\ln\Lambda=20$. Since the heat exchange between the protons and the electrons is proportional to the plasma number density, $n$ (where the plasma is assumed to be quasi-neutral), the thermal coupling between the two species is only important close to the Sun, and becomes negligible at larger distances as the density drops off and the plasma becomes collisionless. The electron energy equation, Eq. (\ref{eq:electrons}), should also include field-aligned thermal conduction, denoted here as $\bfm{q}_e$, and given by the Spitzer form:
\begin{equation}
\bfm{q_e}=-\kappa T_e^{5/2}\frac{\bfm{B}\bfm{B}}{B^2}\bfm{\cdot \nabla} T_e,
\end{equation}
with $\kappa = 9.2\times 10^{-12} W m^{-1} K^{-7/2}$ (calculated by assuming a uniform $ln\Lambda=20$, as before). 

%\section{Wave Dissipation}\label{S:dissip}
\subsection{Turbulent Wave Dissipation}\label{S:dissip}
The coupling of the wave field to the background MHD plasma, as described by equations (\ref{eq:mom}) and (\ref{eq:wavekinetic}) -  (\ref{eq:electrons}), allows us to account for the conversion of wave energy into plasma thermal energy. However, these equations do not explicitly describe the dissipation mechanism itself. In order to complete our description and close the set of equations, we must specify the total wave energy density dissipation rate, $Q_w$.\par 
The nature of the wave dissipation mechanism depends on the local conditions of the plasma in which the waves propagate. In the chromosphere, the plasma is partially ionized and Alfv\'en waves are damped due to finite resistivity / magnetic diffusion effects (c.f. \citet{depontieu2001}). By setting the model's inner boundary at the top of the chromosphere, we can reasonably avoid treating these effects, and only describe the dissipation due to turbulent cascade of Alfv{\'e}n wave energy in the fully ionized corona, where most of the heating takes place. In order to calculate the dissipation rate for any arbitrary magnetic field topology, we apply the unified approach we presented in \citet{sokolov2013}, where open and closed field regions are treated on the same footing. The generalized dissipation term will be discussed below.\par
\subsubsection{Dissipation due to Counter-Propagating Waves} 
The non-linear interaction of Alfv{\'e}nic perturbations can be directly derived  from the MHD equations, by separating the magnetic field and velocity vectors into a background component and a turbulent perturbation component, $\bfm{B}=\tilde{\bfm{B}}+\delta\bfm{B}$ and $\bfm{u}=\tilde{\bfm{u}}+\delta\bfm{u}$. The wave energy densities, $w^\pm$, are related to these perturbations by $w^\pm = \rho |\delta\bfm{u}|^2 = |\delta\bfm{B}|^2 /\mu_0$ (which follows from the equipartition of kinetic and thermal energies of Alfv{\'e}n waves). Substituting these into Eqs. (\ref{eq:mass}) - (\ref{eq:induction}) results in several terms which are second order in the perturbation, essentially describing the evolution of the turbulent energy field due to non-linear wave-wave interactions. The basic derivation can be found in \citet{sokolov2013}. Here we only briefly mention that the dissipation rate due to turbulent cascade will be proportional to the term $\bfm{\nabla\cdot}(\bfm{z}_\mp w^\pm)$, where $\bfm{z}_\pm$ are the Els\"{a}sser variables, defined as $\bfm{z}_\pm= \delta \bfm{u} \pm \delta \bfm{B}/\sqrt{\mu_0 \rho}$. The Els\"{a}sser variables are related to the wave energy densities $w^\pm=\rho z_\pm^2/4$. We can approximate the energy density dissipation rate due to a turbulent cascade as:
\begin{equation}\label{eq:dissip}
Q_w^\pm =\frac{1}{L_\perp}z_\mp w^\pm=\frac{2}{L_\perp}\sqrt\frac{w^\mp}{\rho}w^\pm.
\end{equation}
Here $L_\perp$ is a length scale associated with the transverse correlation length of the turbulent field. Following \citet{hollweg1986}, $L_\perp$ is assumed to be proportional to the width of the magnetic flux tube, which implies that $L_{\perp}\propto 1/\sqrt{B}$. The total dissipation rate (and therefore the heating rate) can be found by summing the contributions from both waves, $Q_w=Q_w^+ + Q_w^-$. Thus the total dissipation rate for counter-propagating waves is given by:
\begin{equation}\label{eq:dissiptot}
Q_w=\frac{1}{L_\perp\sqrt{\rho}}(w^+\sqrt{w^-}+w^-\sqrt{w^+}),
\end{equation}
where the factor of 2 was absorbed into $L_{\perp}$ for simplicity. The value of $L_{\perp}$ is not well-known, but can be constrained by comparison to observations (see Section (\ref{S:free}) for more details).\par 
It is useful to compare Eq. (\ref{eq:dissiptot}) to the phenomenological dissipation term appearing in previous works developed in the framework of Els\"{a}sser variables (c.f. \citet{hossain1995}, \citet{zhou1990}, \citet{matthaeus1999}, \citet{dmitruk2001, dmitruk2002}, \citet{cranmer2007}, \citet{chandran2009}), wherein the total energy density dissipation rate is given by:
\begin{equation}\label{eq:QCranmer}
Q_w^{*} = \rho \epsilon_{turb}\frac{z_+^2 z_- + z_-^2 z_+}{4L_\perp}=\frac{\epsilon_{turb}}{L_\perp\sqrt{\rho}}(w^+\sqrt{w^-}+w^-\sqrt{w^+}),
\end{equation}
where $\epsilon_{turb}$ is a constant specifying the efficiency of the turbulent dissipation (i.e. the ratio of dissipated energy to the injected energy). In the last step we have written this expression in terms of the wave energy densities and absorbed a factor of 2 into $L_{\perp}$ for consistency with Eq. (\ref{eq:dissiptot}). It can be easily seen that Eq. (\ref{eq:QCranmer}) is almost identical to the total dissipation rate given by Eq. (\ref{eq:dissiptot}), differing only by the additional factor of $\epsilon_{turb}$. In \citet{dmitruk2003}, it was shown that $\epsilon_{turb}$  will in general depend on the relative magnitude of the Alfv{\'e}n travel time, $\tau_A$, and the reflection time scale, $\tau_R$  (as well as on the time scales associated with the driving wave field). Simply stated, the efficiency of turbulent heating depends on 
whether the cascade process had sufficient time to develop and heat the plasma before the wave energy is propagated away.  \citet{dmitruk2003} found that $\epsilon_{turb}$ can take values between $13 - 60\%$ for a set of numerical simulations, where the efficiency increases as the reflections become stronger. In the limit of a fully developed cascade where $\tau_{R}<<\tau_{A}$, the efficiency, $\epsilon_{turb}$, will approach unity, and therefore $Q_w \approx Q_w^{*}$. Thus the dissipation rate presented in Eq. (\ref{eq:dissiptot}) is consistent with that derived in previous works, if a fully-developed turbulent cascade is assumed. Even if this assumption is relaxed,  Eq. (\ref{eq:dissiptot})  will only differ by a factor of order unity from Eq. (\ref{eq:QCranmer}).
\subsubsection{Dissipation due to Wave Reflections in Open Flux Tubes}
On closed-field lines, waves of opposite polarities are launched from the two foot-points, and Eq. (\ref{eq:dissiptot}) gives an adequate description. On the other hand, if only one wave polarity is present, $Q_w$ will reduce to zero. In the real solar atmosphere both wave polarities will also be present on open field lines, to some degree, due to reflections. If the local reflection coefficient is given by $C_{refl}$, then the energy density of the reflected wave, $w^{refl}$, is related to the energy density of the outgoing wave, $w^{out}$, by $w^{refl} = C_{refl}^2w^{out}$. 
However, since our model does not explicitly describes reflections, an important distinction has to be made between the theoretical wave energies $w^{out}$, $w^{refl}$ and the model variables $w^\pm$. For a flux tube with $B_r>0$, for example, the variable $w^+$ can be associated with the energy density of the outgoing wave, and we can set $w^+ = w^{out}$. However, we cannot associate the variable $w^{-}$ with $w^{refl}$, since the actual wave reflection was not calculated. In fact, the variable $w^-$ will be equal to zero in this region (up to a round-off error). The opposite will be true in regions where $B_r<0$. In order to properly calculate the dissipation rate in open flux tubes, we must consider a "virtual" reflected wave. This wave will have an energy density equal to $w^{\pm}_* = C_{refl}^2w^{\mp}$, and the energy density dissipation rate of the outgoing wave will then become:
\begin{equation}\label{eq:dissipout}
Q_w^{\pm} =\frac{1}{L_{\perp}\sqrt{\rho}}\sqrt{w^{\mp}_*}w^{\pm}=\frac{1}{L_{\perp}\sqrt{\rho}}C_{refl}\left(w^{\pm}\right)^{\frac{3}{2}}.
\end{equation}
This expression gives the correct energy dissipation rate along open field lines, by taking into account local reflections, without directly simulating the reflections themselves. Note that the above dissipation rate has a similar form as the one derived in \citet{hollweg1986} for open flux tubes, namely $Q_w = (1/L_{\perp}\sqrt{\rho})w^{3/2}$, where $w$ was defined there as the wave energy density of a single polarity. However, the two forms differ by the factor $C_{refl}$, which in the solar corona is estimated to have values between 0.01 and 0.1 (see Section \ref{S:free} for more details). 
\subsubsection{Generalized Wave Dissipation Rate}
The next step is to combine the counter-propagating wave dissipation with the reflected wave dissipation into a single dissipation term that can be applied everywhere. To do so, we note that at any given location either of these mechanisms will be the dominant one, depending on the level of imbalance between the two wave polarities. Thus, we can write:
\begin{equation}\label{eq:dissipfinal}
Q_w^\pm =\frac{1}{L_\perp\sqrt{\rho}}\sqrt{max(w^\mp, C_{refl}^2w^\pm)}w^\pm.
\end{equation}
This form ensures that in regions where both wave polarities have energies within the same order of magnitude (which will occur in closed field line regions), the counter-propagating wave dissipation as it appears in Eq. (\ref{eq:dissiptot}) will be taken into account, while in open field regions or very close to the inner boundary,
Eq. (\ref{eq:dissipfinal}) will reduce to Eq. (\ref{eq:dissipout}).
The advantage of this form is both practical and conceptual. First, the magnetic topology does not have to be determined a-priori in order to "select" a dissipation mechanism (thus making the computational implementation more efficient). More importantly, our form of wave dissipation will cause the distribution of coronal temperatures and wind speeds to emerge naturally and self-consistently with the magnetic topology. This can be understood as follows. The dissipation rate in closed field lines will be larger than that in coronal holes, due to the presence of two wave polarities. This will result in higher heating rates in helmet streamers compared to coronal holes. At the same time, the lower heating rates within coronal holes will lead to more wave energy being available to accelerate the plasma, resulting in a faster solar wind. Another consequence of Eq. (\ref{eq:dissipfinal}) is that the heating rate in active regions will be higher than in the quiet Sun. To see this, we recall that the transverse correlation length, $L_\perp$, is inversely proportional to $\sqrt{B}$. Consequently, regions with higher magnetic field will have a shorter dissipation length scale, and larger dissipation rates. All in all, our choice of the wave dissipation term is capable of self-consistently reproducing the large scale properties of the solar corona and solar wind without invoking geometric heating functions. The only free parameters in this description are the transverse correlation length, $L_\perp$,and the reflection coefficient, $C_{refl}$. We will discuss how we can constrain their numerical values in Section (\ref{S:free}).

\subsection{Constraints on Adjustable Input Parameters}\label{S:free} 
The adjustable input parameters used in this model are:\par 
$\bullet$ The transverse correlation length at the inner boundary, $L_{\perp,0}$.\par
$\bullet$ The pseudo-reflection coefficient, $C_{refl}$, which is assumed to be uniform everywhere.\par
$\bullet$ The Alfv{\'e}n waves Poynting flux at the inner boundary.\par 
\subsubsection{Transverse Correlation Length, $L_{\perp,0}$}
This parameter is used to determine the local correlation length, $L_\perp$, everywhere in the computational domain. Following \citet{hollweg1986}, we assume that $L_\perp$ is proportional to the width of the magnetic flux tube. Due to the conservation of magnetic flux the local correlation length will scale with the magnetic field as $L_{\perp}= L_{\perp,0}/\sqrt{B[T]}$ (where $B[T]$ stands for $B$ measured in units of Tesla).
In the present work we found that a value of $L_{\perp,0}=25$ $km$ gives the proper heating and acceleration rates for solar minimum, by comparing model results to observations. We next compare this value with that employed in other models, and give a general constraint on the choice of $L_{\perp,0}$ for future applications. \citet{hollweg1986}, which solved the problem for coronal hole flux tubes, has estimated $L_{\perp,0}$ to be $75$ $km$. Note, however, that in this latter work, the reflection coefficient was in effect absorbed into the dissipation length. Thus in comparing our formulation (as in Eq. (\ref{eq:dissipfinal})) to the Hollweg one, we have $C_{refl}/L_{\perp,0} = 1/75$ $km$, so the discrepancy between the values used in this current work  \citet{hollweg1986} is not meaningful. Other models which incorporated a more sophisticated description of the turbulent field were found to be in good agreement with observations using values such as $28.76$ $km$ \citep{cranmer2007} and $115.5$ $km$  \citep{cranmer2005}. More recently, \citet{cranmer2010} has determined $L_{\perp,0}$ to be around $60$ $km$, while \citet{sokolov2013} estimated that the correlation length should be in the range $20 - 100$ $km$, which more or less overlaps the values of previous works. Thus we conclude that our choice of the dissipation length is within the range of previous works. A smaller dissipation length will lead to excessive heating close to the inner boundary, and less wave energy will be available for solar wind acceleration farther away.
\subsubsection{Pseudo-Reflection Coefficient}
The reflection coefficient of Alfv{\'e}n waves traveling in an inhomogeneous medium will depend both on the wave frequency, as well as on the gradients in the density and magnetic field. In the present work, we consider, as a first approximation, a uniform reflection coefficient, which can be thought of as an average over the spectrum and over the spatial variance of the plasma. In order to be consistent with previous estimations of the reflection coefficient (c.f. \citet{velli1993}), we restrict $C_{refl}$ to take values between $0.01$ and $0.1$. The actual value chosen for specific simulations will appear in the relevant sections. A more realistic description of the corona should be based on a self-consistent and therefor spatially-varying reflection coefficient. The assumption of a uniform reflection coefficient can be justified for a global model if one compares the predicted and observed of Alfv{\'e}n wave amplitude in the heliosphere (see \ref{S:dissip_results}), as well as compare the resultant solution to that obtained from a more rigorous treatment of wave reflections. Such a comparison will be presented in Landi et al. (2013, in preparation).
\subsubsection{Poynting Flux}
The Poynting flux from the chromosphere to the corona determines the energy input to the model.
Detailed observation of perturbations in the chromosphere have suggested they are likely Alfv\'enic in nature, and their power spectrum was estimated \citep{depontieu2007,mcintosh2012}. The Poynting flux associated with Alfv\'en waves is given by:
\begin{equation}
\bfm{S}=(\bfm{u}\pm \bfm{V}_A)\rho \overline{\delta u},
\end{equation}
where we define the time averaged velocity amplitude as $\overline{\delta u} = \sqrt{<\delta\bfm{u}^2>}$. At an inner boundary at the top of the chromosphere, the flow speed is negligible and we may set $\bfm{u}=0$. The absolute value of the Poynting flux along the magnetic field is then:
\begin{equation}\label{eq:Poynting}
S_{||}= \sqrt{\frac{\rho}{\mu_0}}B\overline{\delta u},
\end{equation}
where $B=|\bfm{B}|$. The numerical value of $S_{||}$ at each point on the inner boundary is therefore completely specified if the plasma density, wave amplitude and magnetic field magnitude are known. The local magnetic field at the inner boundary is derived from either a synoptic magnetic map or an imposed dipole field. For lack of similar global observations of chromospheric Alfv\'en waves, we set $\overline{\delta u}$ to be uniform at the inner boundary, and constrain its value using the observations reported in \citet{depontieu2007}, which found $\overline{\delta {u}}$  to be in the range of $12-15$ $km$ $s^{-1}$ at the altitude where the number density is $n=2\times 10^16$ $m^{-3}$.\par 
We wish to examine the validity of our approximation by comparing the resultant Poynting flux to other models and observational constraints. Inserting the values given above into Eq. (\ref{eq:Poynting}), we get: $S_{||}\approx 0.74 - 1.16\times 10^2 B$ $W m^{-2}$.
Note that the lower limit agrees well with the Poynting flux assumed in the \citet{suzuki2006} model ($0.7\times 10^2 B$ $W m^{-2}$), while the upper limit is comparable to that employed by the unsigned flux heating model \citep{abbett2007}, estimated at $S = 1.1\times 10^2 B$ $W m^{-2}$. A more comprehensive comparison, including to empirical heating models, can be found in our earlier paper, \citet{sokolov2013}.\par

\section{Numerical Model}\label{S:numerical}
The model is implemented within the Space Weather Modeling Framework (SWMF), and is based on the BATS-R-US code, a versatile, massively parallel MHD code developed at the University of Michigan. Detailed description of the BATS-R-US code and the SWMF can be found in \citet{toth2012} and references therein.
BATS-R-US provides a variety of schemes and solvers designed for finite-volume cell-centered numerical methods. In the present implementation, the model equations are solved by a second-order numerical scheme. We found that best results are achieved for this specific model by using an explicit scheme. However, the heat conduction term in Eq. (\ref{eq:electrons}) requires the calculation of second-order derivatives in space, and may constitute a stiff source term, especially in regions of sharp temperature gradients (which will occur near the inner boundary). This may lead to a significant slowing down of the calculation when solved explicitly. In order to overcome this difficulty, we use operator splitting to first solve the hyperbolic operators and non-stiff source terms using an explicit time step, followed by a step which updates the heat conduction term implicitly.\par
Stability is guaranteed by setting the Courant number at $0.8$ \citep{courant1928}. Although this ensures stability for the hyperbolic terms in the model equations, the inclusion of source/loss terms such as wave dissipation, radiative cooling and heat conduction, may lead to negative thermal and/or wave energies. We therefore must further limit the time step by requiring that the loss accrued during a time-step due to any of these processes, and at any given cell, will not exceed the available energy. This is done automatically at runtime, and separately for each of the thermal and wave energy variables.\par 

\subsection{Computational Grid}\label{S:grid}
The use of the SWMF allows us to separate the solar wind model into two coupled physical components - the Solar Corona (SC) component, and the Inner Heliosphere (IH) component. This allows us to optimize our choice of physics, grid geometry and numerical scheme in each domain. The inner boundary of the SC component is located at the top of the chromosphere (which we set at $r=1R_s$), and the outer boundary can be anywhere in the heliosphere, provided that the flow speed at that distance exceeds the fast magnetosonic speed, in order to allow for outflow boundary conditions. Nominally, we set the outer boundary at $r=24R_s$. The inner boundary of the IH components is set at $r=16R_s$, while the outer boundary is set at a distance of a few AU, depending on the application. The coupling between the two components is performed such that the IH component derives its inner boundary conditions from the overlapping cells in the SC domain. The coupling is performed such that second order accuracy in space is maintained.\par 
The model equations are solved on a three dimensional logically Cartesian spatial grid. The computational cells are organized in a block tree, such that each block is composed of the same cell structure. The SC component uses a spherical grid with a block structure of 6x4x4 cells, corresponding to the number of cells in the $(r,\phi,\theta)$ direction. The IH component uses a Cartesian grid with a block stricture of 4x4x4, corresponding to the number of cells in the $(x,y,z)$ directions. 
The capabilities of the BATS-R-US code also include a solution adaptive mesh refinement (AMR), in which blocks are refined by dividing each block into 8 daughter blocks with the same cell structure. The refinement level of neighboring blocks can differ by up to one level of refinement, such that resolution jumps are limited to a factor of 2 in each direction. For the steady-state solutions presented in this paper, AMR is used to automatically resolve current sheets, as we describe in Section (\ref{S:geometry}). The resulting grid typically has 3 million cells in the SC domain and 10 million cells in the IH domain.\par
\subsubsection{Resolving the Transition Region}\label{S:tr}
\begin{figure} 
%\begin{center} 
%\epsscale{.80}
\plotone{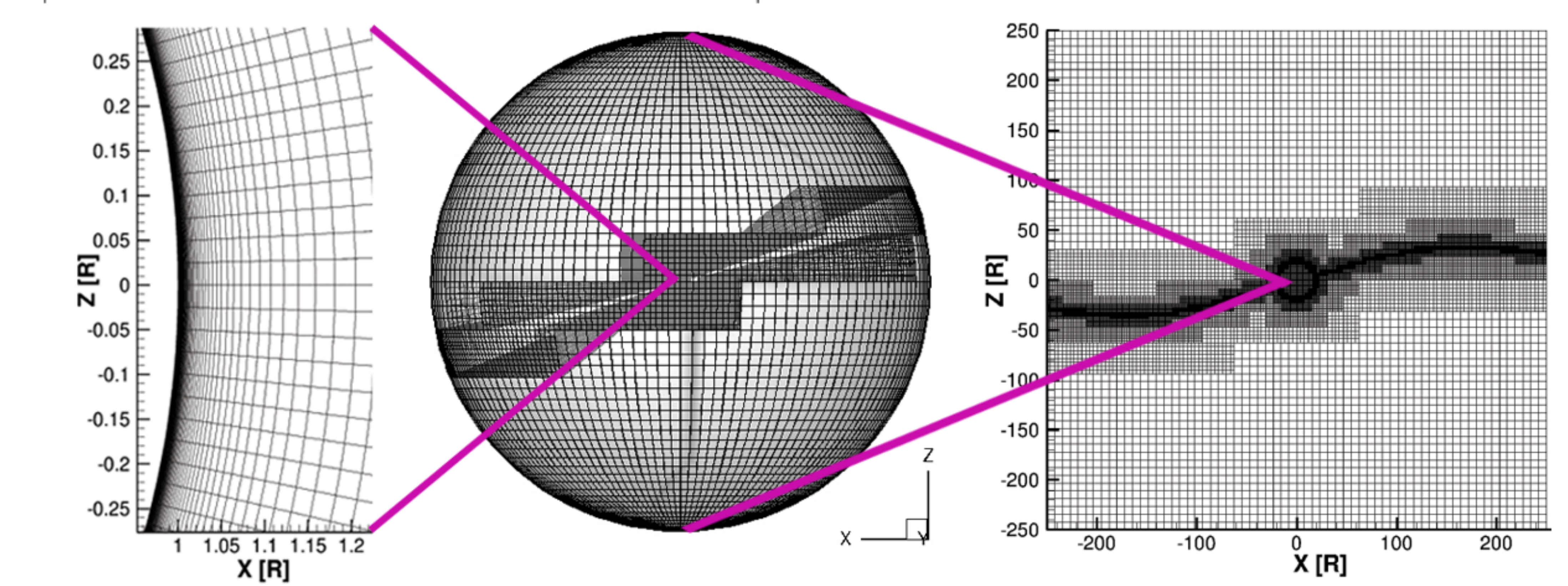} 
\caption{\small \sl The computational grid structure for a pure dipole simulation, with the dipole axis tilted at 15 degrees from the Z axis. Left: The SC (Solar Corona) component grid, near the inner boundary, where the transition region refinement is applied. Center: The entire SC grid, extending up to $24$ $R_s$. Right: The IH (Inner Heliosphere) component grid. In both the SC and IH components, a finer grid is automatically created by AMR due to the presence of the heliospheric current sheet (in blocks where the radial magnetic field changes sign).\label{F:grid}} 
%\end{center} 
\end{figure}

In order to allocate sufficient resolution to the transition region and lower corona, while minimizing the number of cells at larger heliocentric distances, we use non-uniform grid spacings in the radial direction. Building on the work presented in \citet{downs2010}, we construct the radial spacings such that more grid points are concentrated close to the Sun. The magnitude of the radial spacings $\Delta r$ is a smooth function of $\ln(r)$, becoming uniform in $\ln(r)$ beyond $r=1.7R_s$. The resulting grid is depicted in the left panel of Figure \ref{F:grid}. \par 
The smallest radial spacing, occurring near the inner boundary and inside the transition region, is $\Delta r =0.001R_s \approx 700$ $km$. However, the typical length scales of the dynamic processes in the transition region can be as small as a few kilometers. Resolving the transition region to these scales is impractical in the framework of a global model extending to the solar wind. We therefore use the method presented in \citet{lionello2009}, in which the following transformation is applied to the model equations:
\begin{equation}
Q_w \rightarrow Q_w/f \qquad Q_{rad}=Q_{rad}/f \qquad \kappa_0\rightarrow f\kappa_0 \qquad ds\rightarrow fds ,
\end{equation}
where  $ds$ is the path length along a field line and $f$ is a scalar factor given by:
\begin{equation}
f = \left(\frac{T_m}{T_e}\right)^\frac{5}{2} ,
\end{equation}
where $T_m$ is some constant reference temperature, and $T_e$ is the local electron temperature. This transformation essentially rescales the energy equation. For $T_e < T_m$ we will have $f>1$, effectively increasing the characteristic length scale of the processes participating in the energy balance, thus widening the temperature profile in the transition region. We must choose $T_m$ such that the length scale in the transition region will be increased so as to accommodate several grid points. As estimated in \citet{sokolov2013}, this condition will be satisfied for $T_m=220,000K$. We must also require that this transformation will not affect the coronal solution, which is sufficiently resolved, and so the transformation is only applied in the range $T_{0}<T_e<T_m$ where $T_0$ is the temperature at the inner boundary, $T_0=50,000K$. Note that $f$ smoothly approaches unity at $T_e=T_m$, thus ensuring the widened temperature profile at the transition region will smoothly connect to the coronal temperature profile.\par
\begin{figure} 
\begin{center} 
%\epsscale{.80}
\plotone{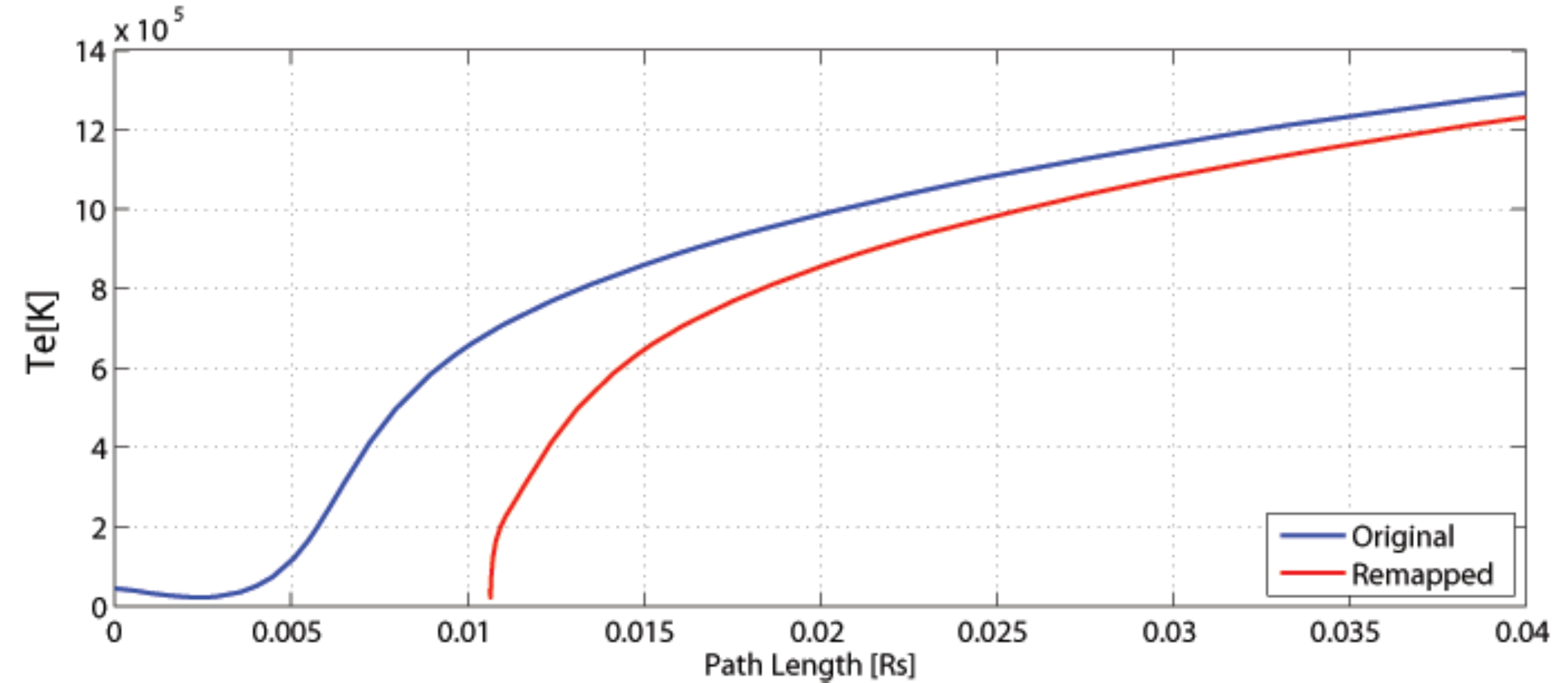} 
\caption{\small \sl Temperature profile taken along a closed field line in the streamer belt, from an ideal dipole simulation. The blue curve shows the modeled profile. The red curve shows the temperature profile after remapping it using the inverse scale transformation.\label{F:remap_streamer}} 
\end{center} 
\end{figure}

 
Although this transformation will not affect the solution in the corona and solar wind, care has to be taken when comparing our model results to observations in the lower corona. In this case we must map modeled profiles back into realistic scales, by applying the inverse transformation. An example of this procedure is given in Figure \ref{F:remap_streamer}, showing the temperature profile along a streamer belt field line in an ideal dipole simulation. The blue curve shows the model result, and the red curve shows the remapped profile. One can see how the modeled temperature profile is gradually compressed by the mapping, restoring the sharp temperature gradient in the transition region. This procedure should be repeated when calculating line-of-sight integrals as well (as is done, for example, when creating synthesized images). In what follows, we will show original model results, without the remapping, unless otherwise specified.

\subsubsection{Other Geometric Considerations}\label{S:geometry}
The spherical nature of the problem makes a spherical grid a natural choice for the SC component. However, the simple spherical grid introduced here will give rise to a singularity along the polar axis, where cell faces touching the pole will have a zero area. This means that fluxes cannot move across the pole. In order to overcome this, we use the super-cell algorithm described in \citet{toth2012}. We apply the super-cell algorithm to a single layer of cells surrounding the pole, from the inner boundary and up to the edge of the SC domain.\par 
In both SC and IH components, we use adaptive mesh refinement in order to resolve current sheets. The criterion for refining a block is whether the radial component of the magnetic field changes sign inside the block. The largest current sheet is the heliospheric current sheet, a thin current layer originating from coronal hole boundaries and extending over the entire heliosphere. Although its topology is wrapped by solar rotation, it remains a rather thin layer throughout the heliosphere. Since cell sizes increase with radial distance in a spherical grid, a Cartesian grid is a more suitable choice for the IH component. The current sheet refinement is excluded from regions with $r< 1.7$ $R_s$, so as to avoid over-refinement in the transition region grid. Figure \ref{F:grid} shows the resulting refinement for the case of a pure dipole that is tilted by 15 degrees from the Z axis.

\subsection{Inner Boundary Conditions}\label{S:bcs}
Synoptic magnetograms of the photospheric magnetic field are routinely obtained by several solar observatories, and their use in global coronal models is widespread. Here, we use synoptic magnetograms to specify the radial component of the magnetic field at the inner boundary.\par 
The temperature and density are assumed to be uniform at the inner boundary. The proton and electron temperatures are set to $T_e=T_p=50,000K$. The particle number density can take values in the range $n=n_e=n_p=2\times 10^{16} \div 2\times 10^{17}$ $m^{-3}$. The mean velocity amplitude of the Alfv{\'e}n waves, $\overline{\delta u}$, is uniform at the inner boundary as well. Under these assumptions, the Poynting flux defined in Eq. (\ref{eq:Poynting}), will vary with the surface magnetic field according to $S_{||} = C_S B$ $W m^{-2}$ where $C_S$ is a constant. As discussed in Section \ref{S:free}, we constrain the wave amplitude to take values in the range $\overline{\delta u} = 12-15$ $km$ $s^{-1}$  at the altitude where the density is $n=2\times 10^{16}$ $m^{-3}$, leading to a Poynting flux per unit magnetic field in the range $C_S = 0.74 - 1.16\times 10^2$ $W m^{-2} G^{-1}$. If the simulation is to start at a lower altitude with higher number density, the Poynting flux at the inner boundary should be increased such that the desired flux is obtained at the altitude where $n=2\times 10^{16}$ $ m^{-3}$.\par 
Once the Poynting flux at each point on the inner boundary is known, we calculate the wave energy density according to $w^\pm = \rho \overline{\delta u}$. At each location on the inner boundary, we use the polarity of the magnetic field to determine which wave mode carries the Poynting flux, such that it is only carried by an outgoing wave. The energy density of the in-going wave is set to zero, so that if any in-going wave reaches the inner boundary (as can occur in closed magnetic loops) then it will be perfectly absorbed.\par 
The radial bulk speed at the solar surface is theoretically zero. However, this implies a null mass flux coming from the inner boundary, and can create unwanted artifacts in the solution. We therefore avoid explicitly specifying the velocity at the inner boundary. Rather, we require a zero electric field, which in the frozen-in regime implies that $\bfm{u}||\bfm{B}$. We thus simply impose field-aligned flow at the inner boundary. The resulting solutions show that this choice leads to very small bulk speeds close to the surface (up to a few kilometers per second), which are later accelerated as expected.

\section{Model Results for Idealized Magnetic Fields}\label{S:ideal}
Ideal cases with simple magnetic topology will help us test the model and gain physical insight into the resulting steady-state solutions. For this purpose, we assume the Sun's intrinsic magnetic field is an ideal dipole field, with a polar field strength of $5.6$ $G$ (which is comparable to the observed polar field during solar minimum). The idealized field is used to define the radial magnetic field at the inner boundary, and the total magnetic field is allowed to evolve self-consistently. 
\subsection{Coronal and Solar Wind Structure}\label{S:results_dipole}
\begin{figure} 
%\begin{center} 
%\epsscale{.50}
\plotone{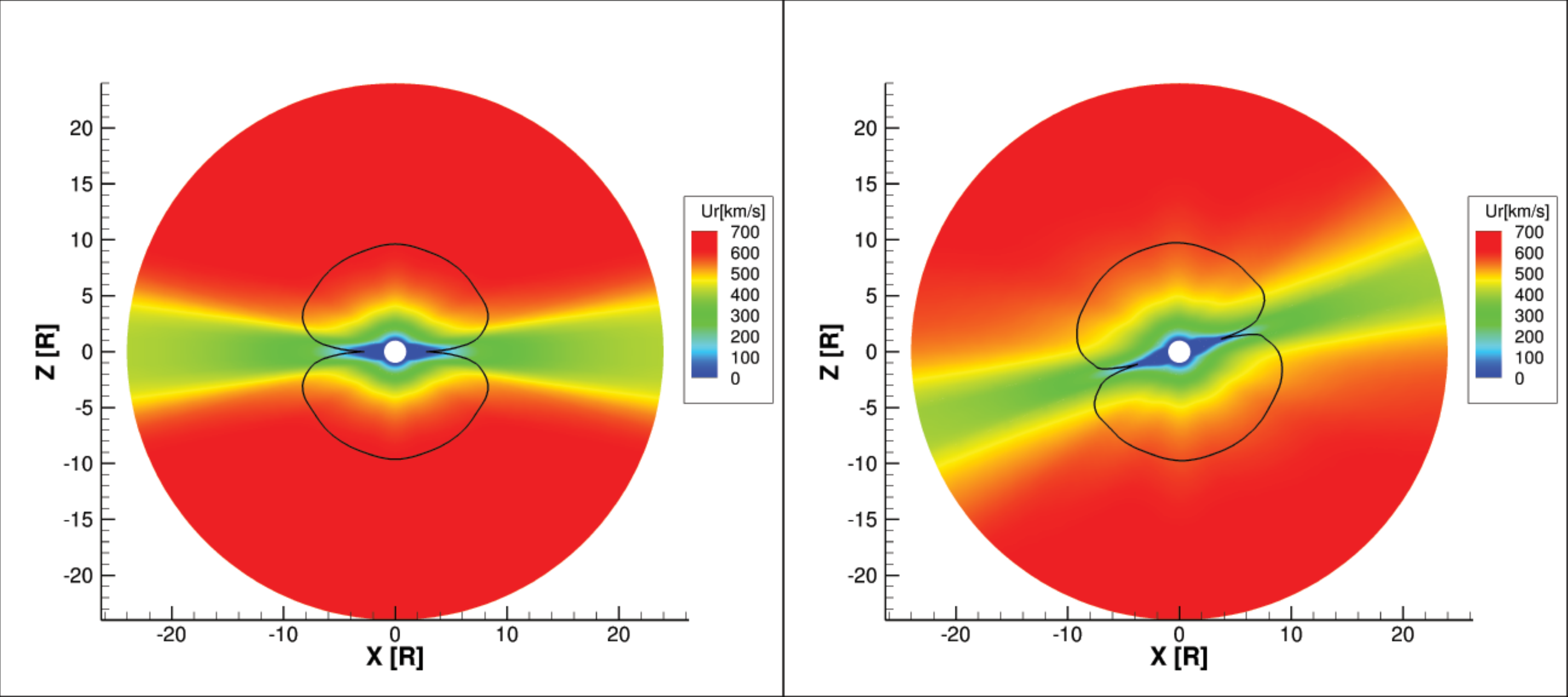} 
\caption{\small \sl Radial velocity in a meridional plane for a two-temperature, ideal dipole simulation. The black curve shows the location of the Alfvn{\'e}nic surface. Left: dipole axis aligned with solar rotation (Z) axis. Right: dipole axis tilted by 15 degrees with respect to the rotation axis. \label{F:dipole_tilt_UrVa}} 
%\end{center} 
\end{figure}

Figure \ref{F:dipole_tilt_UrVa} shows the distribution of radial speeds in the meridional plane up to $24R_s$, taken from steady-state solutions (in a co-rotating frame) of ideal dipole fields. In the left panel the dipole axis is aligned with the solar rotation axis (Z-axis) while in the right panel the dipole axis is tilted by 15 degrees with respect to the Z-axis. The black curve in each panel shows the location of the Alfv{\'e}nic surface, where $u_r = V_{A,r}$. As can be seen, the model produces a velocity distribution of fast and slow solar wind flows. The aperture of the slow solar wind in about 20 degrees from the equatorial plane. The location of the Alfv{\'e}nic surface, at about $8R_s$, is consistent with previous studies.\par

\begin{figure} 
%\begin{center} 
%\epsscale{.70}
\plotone{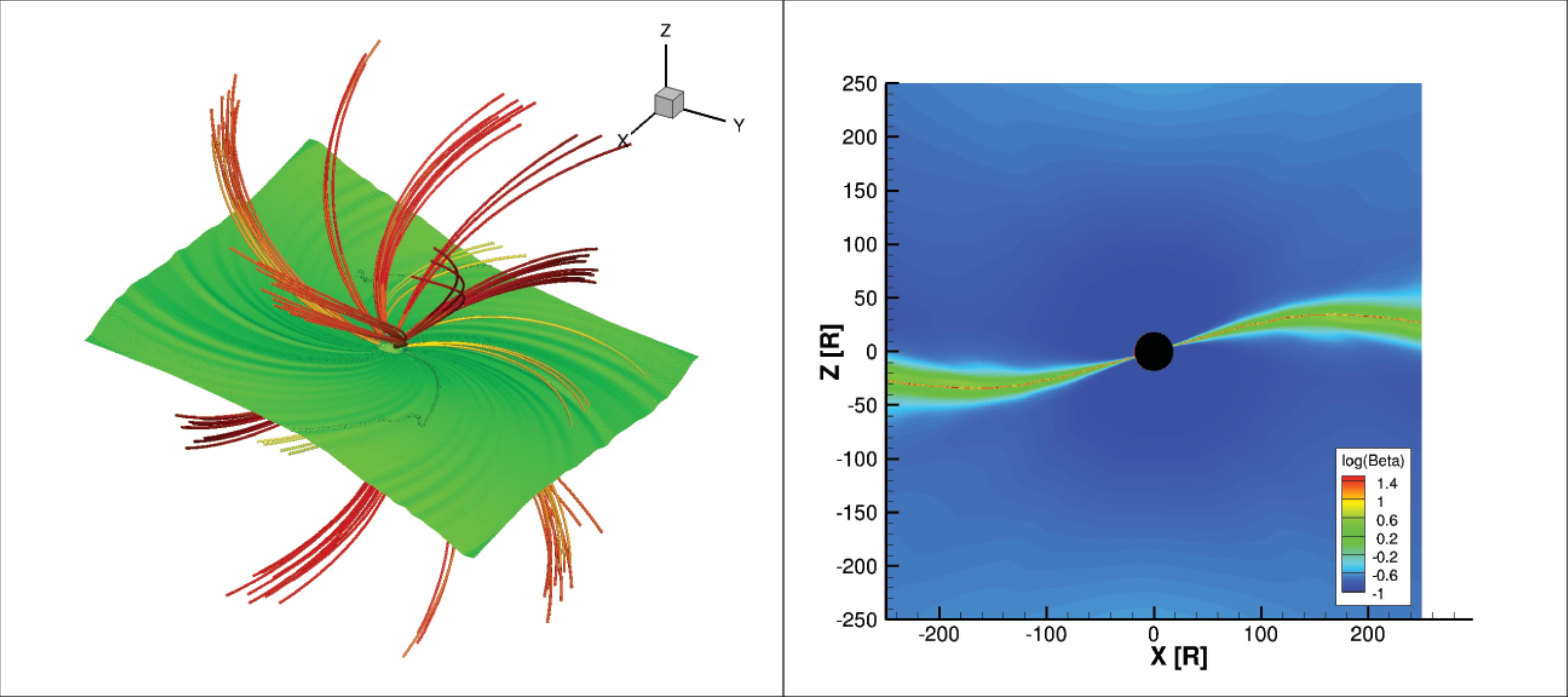} 
\caption{\small \sl Results of a the tilted dipole simulation in the inner heliosphere, up to $250$ $R_s$. Left: 3D structure. Green surface shows the location of the current sheet (where $B_r=0$). Stream lines show the magnetic field, colored by the radial speed (using the same color scale as in Figure \ref{F:dipole_tilt_UrVa}). Right: Plasma beta in the y=0 plane.\label{F:IH_tilt_3D_beta}} 
%\end{center} 
\end{figure} 

As mentioned in Section (\ref{S:geometry}), the singularity at the Z axis of the spherical grid may constitute a numerical challenge, since numerical fluxes are inhibited there and a special treatment of the pole is required. Comparing the cases of the tilted and non-tilted dipole, we have verified that the model produces the expected results even when the symmetry axis of the problem is not aligned with the symmetry axis of the grid. No numerical artifacts seem to be created by the pole singularity.\par 
In the non-tilted dipole case, the problem is azimuthally symmetric. However, when there is a tilt angle between the rotation axis and magnetic axis, the heliospheric current sheet will warp and bend, producing the well-known "Ballerina skirt" further away from the Sun. Figure \ref{F:IH_tilt_3D_beta} shows the steady state solution for the tilted dipole case, up to a heliocentric distance of $250R_s$. The left panel shows the three-dimensional structure of the current sheet (green surface), and the magnetic field lines (colored by radial speed). The right panel shows the plasma beta (ratio of thermal to magnetic pressures). The region of high plasma beta (red) signifies a null magnetic field. This figure demonstrates that the heliospheric current sheet remains thin throughout the simulation domain.

\subsection{Two-Temperature Effects}
\begin{figure} 
%\begin{center} 
%\epsscale{1.80}
\plotone{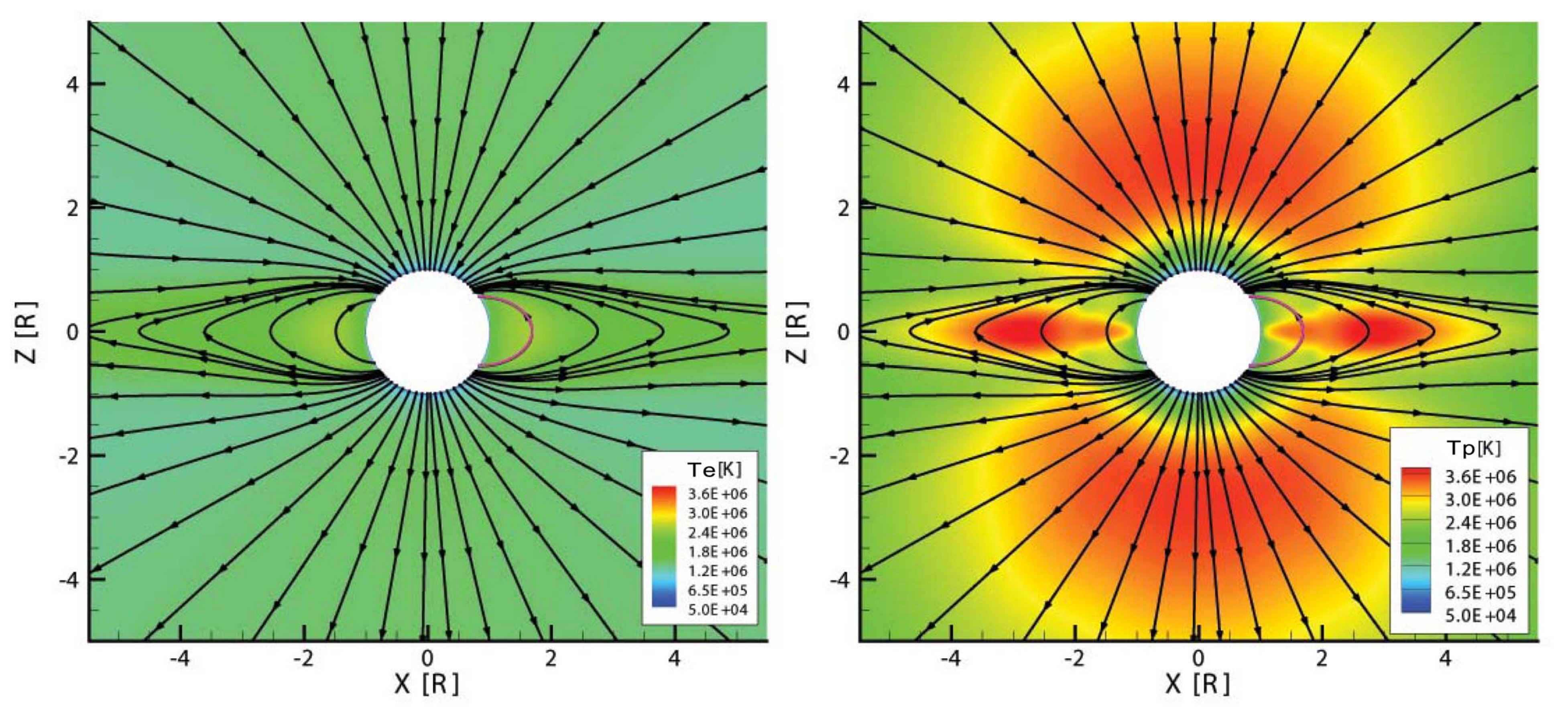} 
\caption{\small \sl Electron temperature (left) and proton temperature (right)  in a meridional plane for an ideal dipole simulation. The black curves show the magnetic field. The purple curve denotes the closed field line used for extracting the data used in figures (\ref{F:loop_teti}) and (\ref{F:loop_waves}).\label{F:dipole_teti}} 
%\end{center} 
\end{figure}

Figure \ref{F:dipole_teti} shows the electron (left panel) and proton (right panel) temperature distribution in a meridional plane. This result demonstrates the combined effects of electron heat conduction and electron-proton thermal decoupling. First, the field-aligned electron heat conduction causes the electron temperature to be almost uniform along closed magnetic field lines. For protons, a clear maximum occurs at the tip of the helmet streamer, where wave dissipation due to counter-propagating waves is largest (see below). Due to the low coronal density, the second term on the right hand side of Eqs. (\ref{eq:protons})-(\ref{eq:electrons}), which gives the electron-proton thermal coupling, becomes negligible at these altitudes. In the absence of a mechanism for the protons to lose their energy, the proton thermal energy remains "trapped" locally. Overall, the protons are about two times hotter than the electrons. This can be understood as follows. Electrons can efficiently conduct excessive heat from the hot corona down to the much cooler transition region and chromosphere, where  the radiative cooling rate is considerably higher due to high plasma densities and low temperatures. Since we assume the radiated energy does not interact with the plasma (which is a reasonable approximation for the corona), the transition region can be viewed as a heat sink for electrons.  At lower altitudes this mechanism also cools the protons due to thermal coupling between the two species, but this process becomes inefficient above the transition region.\par
\begin{figure} 
%\begin{center} 
%\epsscale{.50}
\plotone{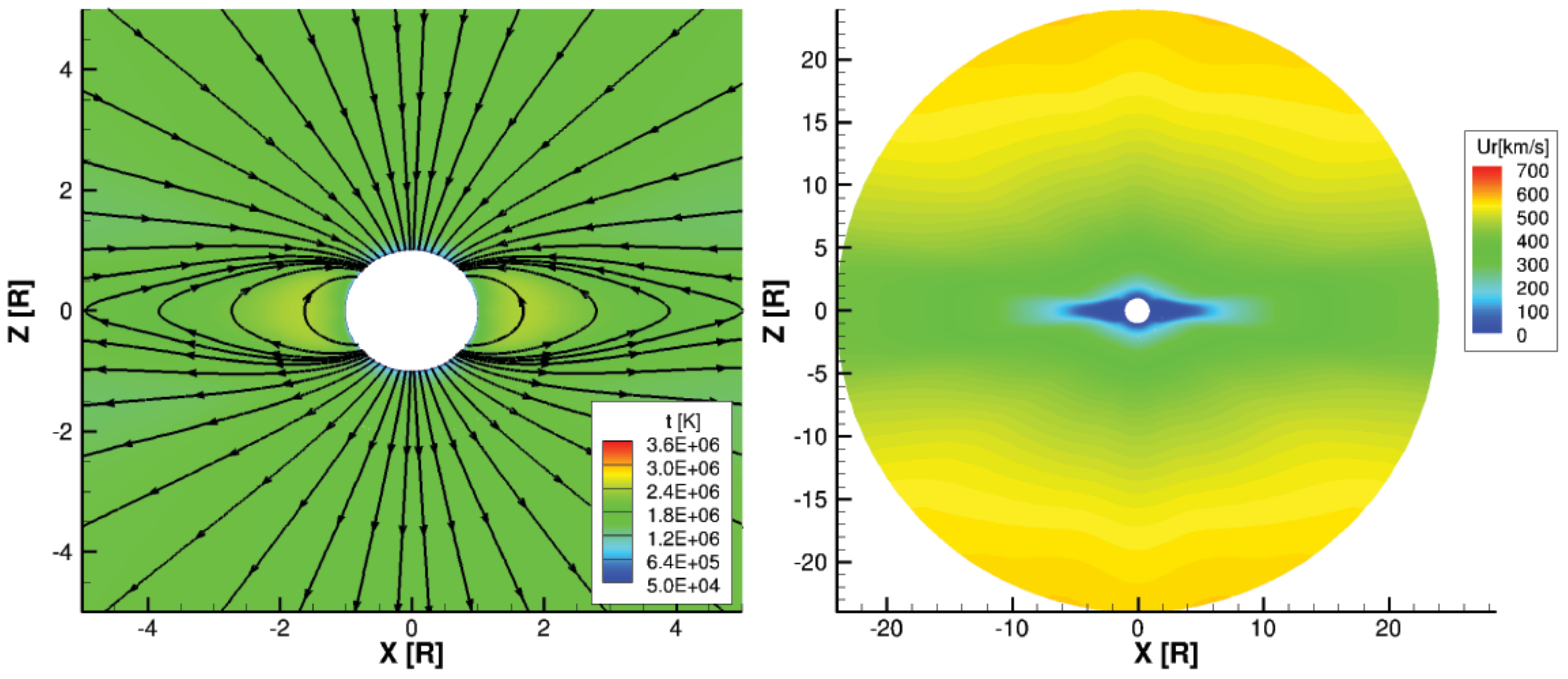} 
\caption{\small \sl  Steady-state solution  in a meridional plane for the single-temperature, ideal dipole simulation. Left: Radial speed and magnetic field lines. Right: plasma temperature.\label{F:1T_dipole_Ur_T}} 
%\end{center} 
\end{figure}

The importance of a two-temperature description can be further demonstrated if we compare the above result to that obtained in a single temperature simulation. This is achieved by setting $p_p=p_e$ in Eqs. (\ref{eq:mom}), (\ref{eq:protons}), and (\ref{eq:electrons}). All other free parameters are kept the same as the two-temperature simulation. Figure \ref{F:1T_dipole_Ur_T} shows the resulting velocity field (left) and plasma temperature (right), in a meridional plane. One can see that in the single-temperature case, the corona is cooler and the solar wind is slower than in the two-temperature case, even though the Poynting flux injected into the system is the same. A single temperature description is equivalent to the assumption that the electrons and protons are in thermodynamic equilibrium, so that wave dissipation and heat conduction affect the plasma as a whole.  In the absence of electron-proton decoupling, less thermal energy can be retained by the protons. This causes more thermal energy to be removed from the system by heat conduction and subsequent radiative cooling. The resulting steady state must therefore be less energetic as a whole for a single-temperature case.\par 
We conclude that a two-temperature description is more realistic than a single-temperature one. The effects of decoupled protons may become more important when describing solar eruptions, where the ejecta can be magnetically connected to the Sun, allowing for thermal energy to flow back to the Sun, thus producing unrealistic shock structures. This is further discussed in \citet{jin2013}.
 
\begin{figure} 
\begin{center} 
\epsscale{0.8}
\plotone{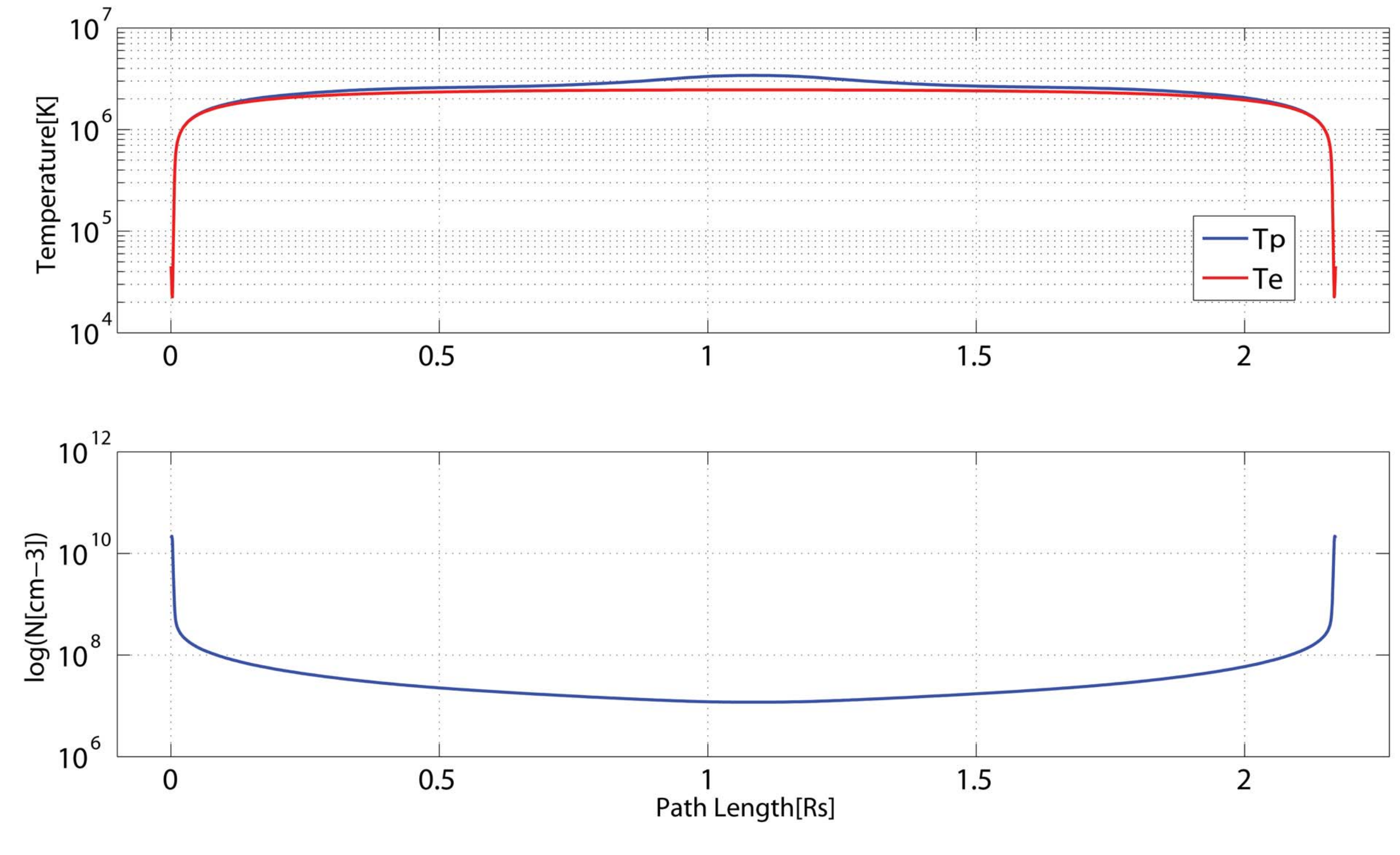} 
\caption{\small \sl  Plasma  properties extracted along a loop in the streamer belt of an ideal dipole solution. Top: electron and proton temperatures. Bottom: density. Data was extracted from the loop shown in purple in figure (\ref{F:dipole_teti}).\label{F:loop_teti}} 
\end{center} 
\end{figure}

\begin{figure} 
%\begin{center} 
%\epsscale{1.0}
\plotone{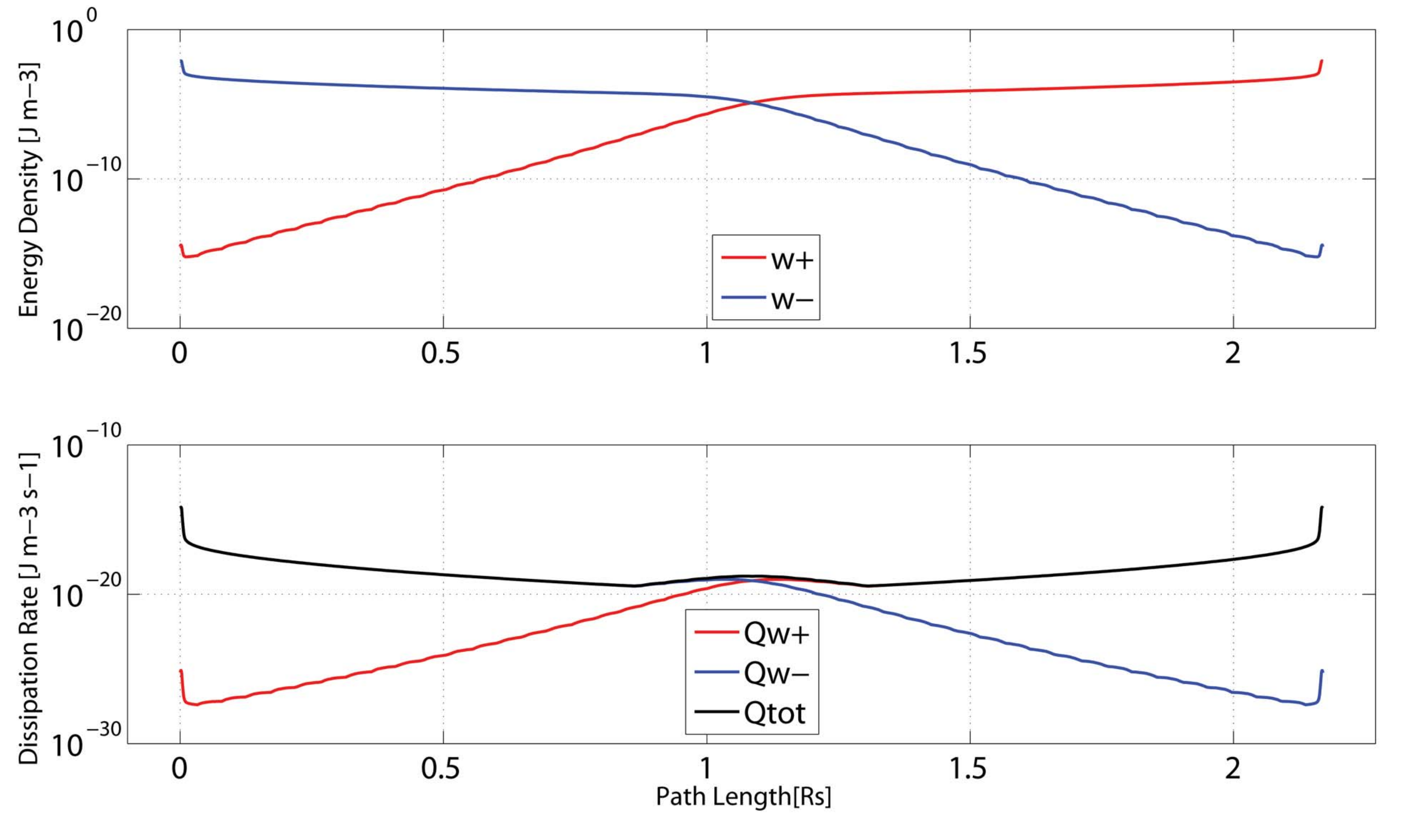} 
\caption{\small \sl Wave energy densities (top) and energy density dissipation rates (bottom) for both wave polarities, extracted along a loop in the streamer belt of an ideal dipole solution, shown as the purple field line in figure (\ref{F:dipole_teti}).\label{F:loop_waves}} 
%\end{center} 
\end{figure} 

\begin{figure} 
%\begin{center} 
\epsscale{0.9}
\plotone{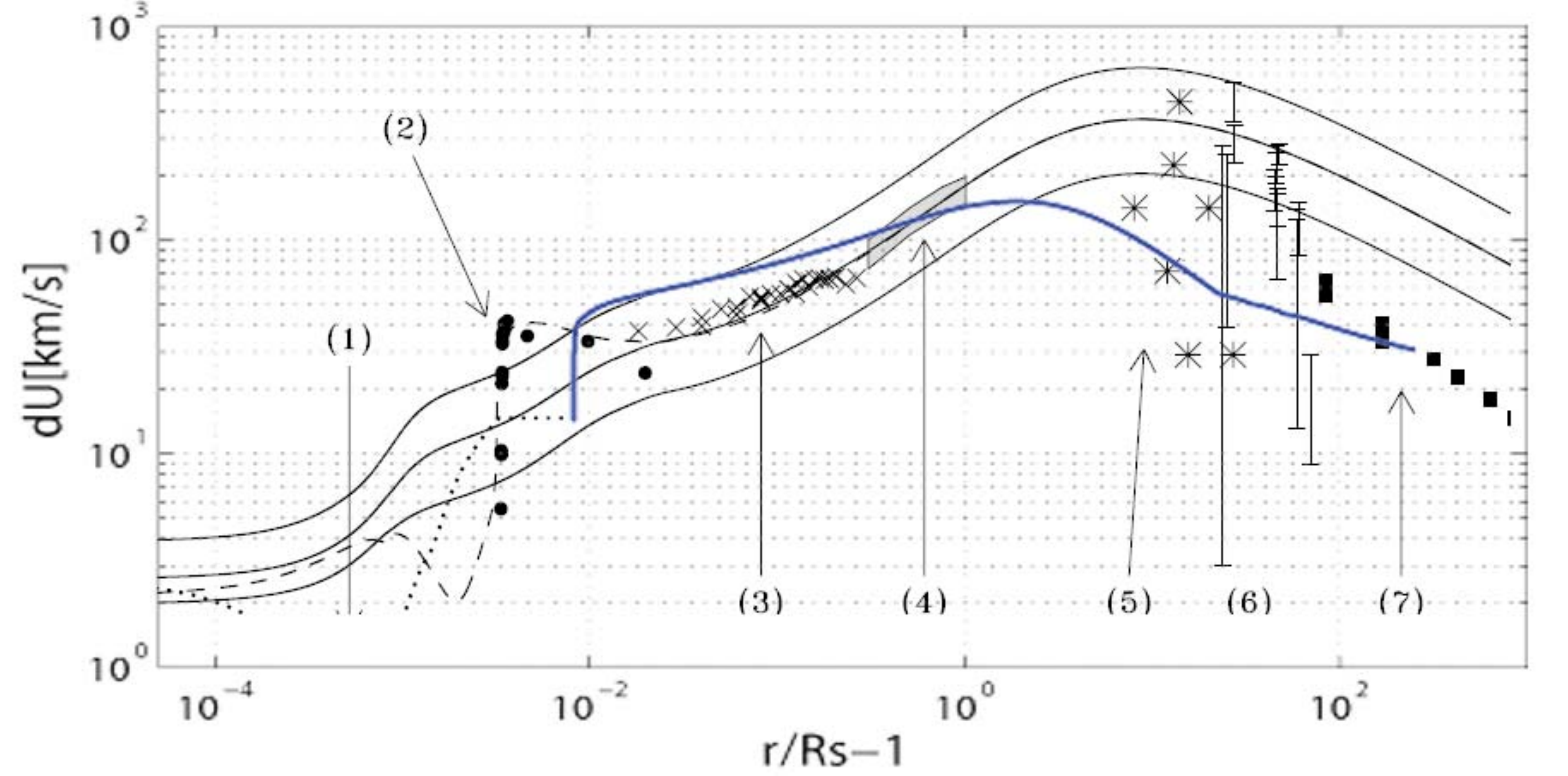} 
\caption{\small \sl Velocity perturbation vs. heliocentric distance from model results and observations. The figure shows AWSoM model results (blue curve) overlaid on top of figure 9 from \citet{cranmer2005}. Black symbols represent observed values, while the black solid curves show the \citet{cranmer2005} model results. The AWSoM results were extracted along a polar coronal hole field line, for an ideal dipole simulation. The numbers (1) - (7) indicate observation sources, see \citet{cranmer2005}.\label{F:polar_dU}} 
%\end{center} 
\end{figure} 

\subsection{The Role of Wave Dissipation}\label{S:dissip_results}
The AWSoM model is the first global model to unify the treatment of open and closed field lines. This is a direct result of Eq. (\ref{eq:dissipfinal}), which describes a wave energy dissipation rate that automatically adjusts to the magnetic field topology, allowing either reflected-wave dissipation or counter-propagating wave dissipation to dominate.\par 
 The interplay between the two types of dissipation mechanisms can be best studied by examining the evolution of the wave energy and its coupling to the plasma along typical magnetic structures, like helmet streamers and coronal holes. Figure \ref{F:loop_teti} shows the electron and proton temperature, as well as the plasma density, extracted along a magnetic loop in the helmet streamer (marked by the purple field line in Figure \ref{F:dipole_teti}). We note that our model reproduces sufficiently well the sharp density and temperature gradients known to exist in the transition region. The temperature profile of the electrons (top panel) is almost flat in the corona, while the protons become hotter at the top of the streamer. In order to study in more detail how this peak is created, we must examine the wave energy density and dissipation rates of both wave polarities. These are shown in Figure \ref{F:loop_waves}. The top panel shows the energy densities of the parallel and anti-parallel waves along the same field line. The two wave modes have their maximum energy at opposite foot points of the streamer loop, since only a single wave mode is launched from each point on the inner boundary. One can see that the energy density sharply decreases at the middle of the loop, reaching negligible amounts at the other foot point. The energy density dissipation rate (bottom panel), is largest in the transition region. Above the transition region, the dissipation rate of each wave mode smoothly decreases from its maximal value at its respective foot point due to the reflected wave dissipation term in Eq. (\ref{eq:dissipfinal}). At the top of the loop, the wave energies of the two modes become comparable, and the counter-propagating term kicks in. This produces a local maximum in the total dissipation rate, and the peak in proton temperature.\par 
The electron temperature in the streamer belt is about $70\%$ higher than that in the coronal holes (see Figure \ref{F:dipole_teti}). This can be understood if we notice that wave dissipation rates will be higher in closed-field regions, where two wave modes are injected into a single field line, while in coronal holes dissipation is only due to reflections. As a result, more wave energy will be available in coronal hole flux tubes, enabling higher acceleration rates due to the action of wave pressure. Thus the temperature distribution is closely related to the velocity field distribution. Examining Figure \ref{F:dipole_tilt_UrVa}, we can immediately recognize that the regions of lower temperatures in the coronal holes correspond to the source region of fast solar wind flows, while the hotter streamer is embedded in a region of slow solar wind. Thus, our choice of wave mechanism automatically produces the observed large-scale temperature and velocity structure of the solar corona and wind.\par 
In order to complete the discussion of wave dissipation, and to further justify our proposed turbulent wave dissipation mechanism, we must examine whether the resulting wave field is consistent with observations. Figure \ref{F:polar_dU} shows the amplitude of the velocity perturbation, $\delta\bfm{u}$, associated with the outgoing wave, as a function of radial distance, calculated along a polar coronal hole. At lower altitudes, where $Te<220,000K$, the profile was rescaled in order to compensate for the artificial transition region broadening, as discussed in Section \ref{S:tr}. This profile is qualitatively in good agreement with the observations compiled in \citet{cranmer2005} (see Figures 7 and 15 therein). In particular, the sharp gradient in wave amplitude close to the inner boundary occurs at roughly the same altitude ($10^{-2}R_s$), and reaches a similar magnitude ($~40$ $km$ $s^{-1}$) in both the model and the observations. The second local maximum occurs around $2R_s$, where the wave amplitude reaches $150$ $km$ $s^{-1}$. Finally, the wave amplitude at 1 AU is about $30$ $km$ $s^{-1}$. These fall within the range of observed values. It should be noted that modeled values will be somewhat different in a steady state solution corresponding to a specific Carrington rotation. Since the available observations span several rotations, we regard the steady-state solution of an ideal dipole field as a proxy for a generic solar minimum configuration.

\section{Model-Data Comparison for Solar Minimum}\label{S:minimum}
In order to directly compare our model results with the variety of available observations, we simulate a steady-state solution for Carrington Rotation CR2063 (11/4/2007 - 12/2/2007), which took place during solar minimum. We compare our results to remote observations in the lower corona, as well as in-situ observations in the solar wind.  We can thus test whether the model can simultaneously reproduce observations at these highly different environments, while the entire system is driven only by the rather simple boundary conditions described in Section \ref{S:bcs}. 

\begin{figure} 
%\begin{center} 
\epsscale{1.0}
\plotone{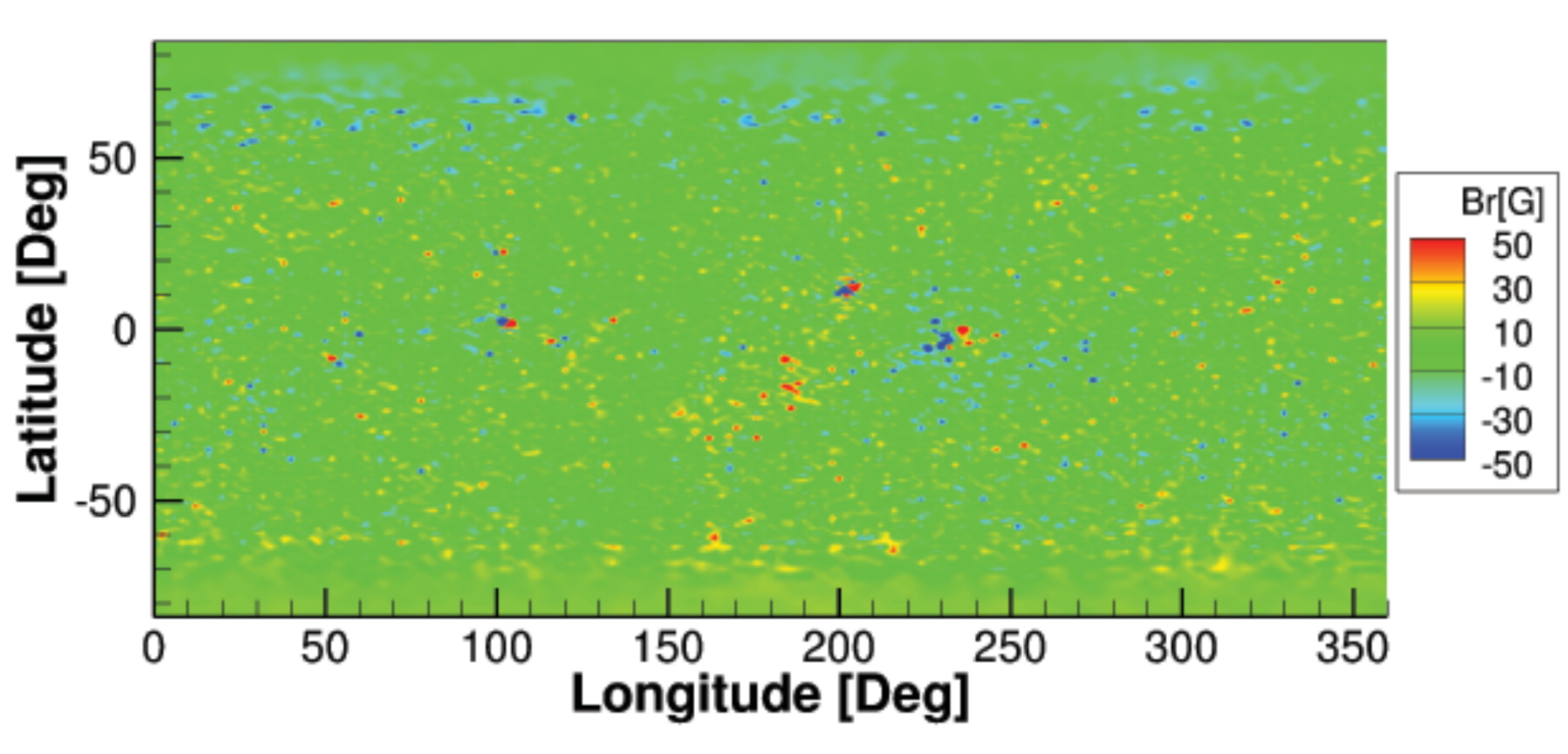} 
\caption{\small \sl Boundary condition for the radial magnetic field for CR2063, obtained from an MDI magnetogram with polar interpolation. Note that the color scale was modified so that the large scale distribution can be seen. However magnetic field intensity can reach up to 500 G in the small regions in the vicinity of active regions.\label{F:2063_Br}} 
%\end{center} 
\end{figure}

\subsection{Model Input and Limitations}
As an input to the model, we set $L_{\perp,0}=25 km/\sqrt{T}$, $C_{refl}=0.06$ and $\overline{\delta u}=15$ $km$ $s^{-1}$. For the magnetic field, we use a line-of-sight synoptic magnetogram obtained  by the Michelson-Doppler Interferometer (MDI) instrument on board the Solar and Heliospheric Observatory (SOHO) spacecraft \citep{scherrer1995}. The magnetogram radial field is used to determine the inner boundary condition for the model.\par 
Line-of-sight magnetograms possess an inherent uncertainty at the polar regions, since the line-of-sight to these regions is almost perpendicular to the radial direction. We therefore use a polar-interpolated synoptic magnetogram, provided by the Solar Oscillations Investigation (SOI) team \citep{sun2011}. Synoptic magnetograms are also known to possess uncertainties in the magnetic field intensity over the entire disk. Several studies have shown that the intensity derived from magnetograms may vary depending on spatial and temporal resolutions, location on the disk, instrument noise and zero-offset bias, and level of solar activity (c.f. \citet{pietarila2012}). Although some of these difficulties are mitigated by proper calibrations, synoptic magnetograms from different instruments may still give different results. MDI data have been found to scale by a factor of 0.6 - 1.4 compared to other instruments \citep{liu2012,pietarila2012}. Since the "true" magnetic field intensity is not known, we increased the magnetogram field for CR2063 by a factor of 2, which we estimated by comparing modeled and observed coronal hole boundaries. We note that without scaling, the magnetogram leads to unrealistically large coronal holes in the model, suggesting that the input field is too weak to contain the plasma and field lines that should be closed are opened up by the plasma flow. The resulting boundary condition for the radial magnetic field is shown in Figure \ref{F:2063_Br}. \par 
It should be noted that the use of synoptic magnetograms, which are collected over a period of a full solar rotation (about 27 days), limits our ability to capture short-lived magnetic structures. The steady-state solution we obtain should therefore be considered as simulating the average state of the system over the period covered by the magnetogram.\par

\subsection{Coronal Density and Temperatures Profiles}
\begin{figure} 
%\begin{center} 
%\epsscale{.70}
\plotone{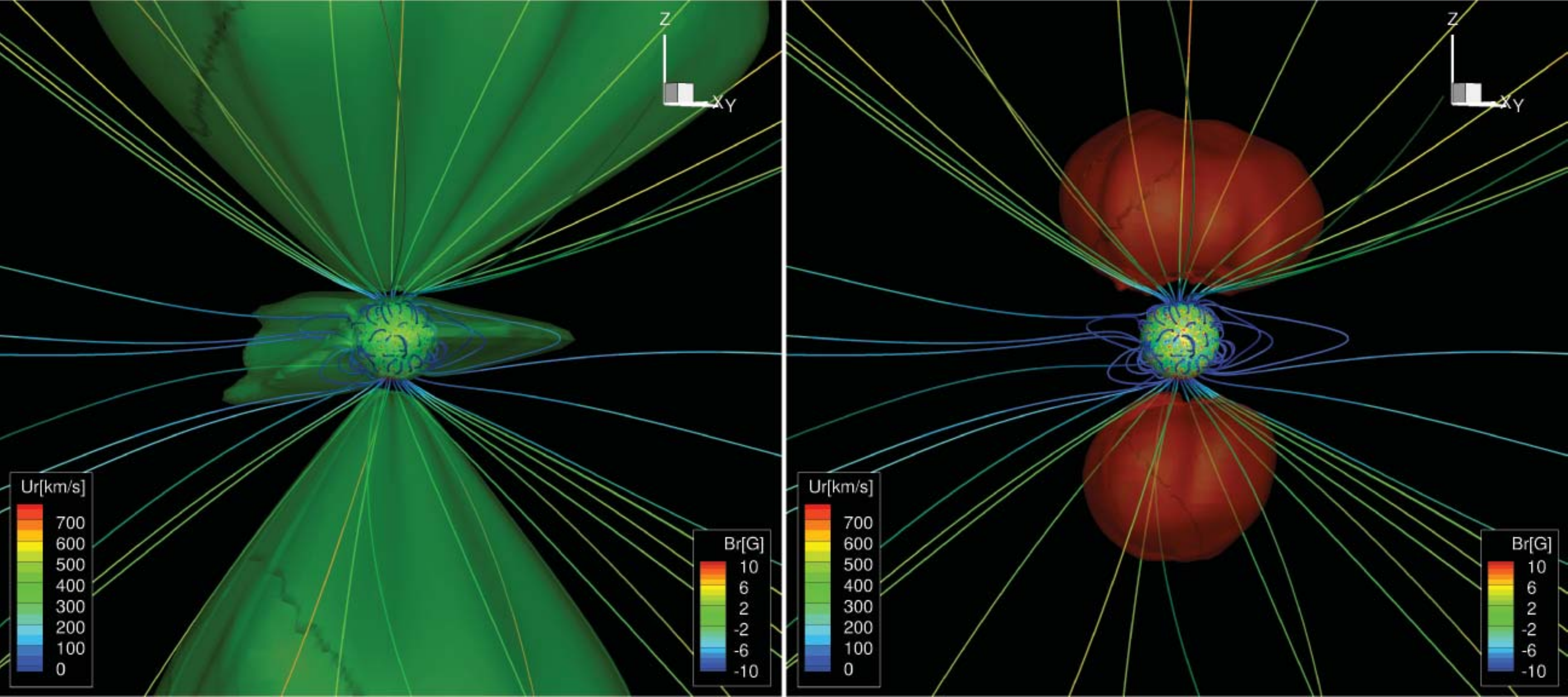} 
\caption{\small \sl Results for CR2063. Solar surface colored by radial magnetic field strength. Field lines are colored by radial speed. The left panel shows a temperature iso-surfaces for electrons at 1.3MK. The right panel shows a temperature iso-surface for protons at 3MK.\label{F:2063_tite}} 
%\end{center} 
\end{figure}

Figure \ref{F:2063_tite} shows the steady-state solution up to $5R_s$. The solar surface is colored by the radial magnetic field. Streamlines denote magnetic field lines, colored by radial speed. Also shown are temperature iso-surfaces for electron and protons (left and right panels, respectively). As expected for a solar minimum configuration, the coronal holes are mostly concentrated around the poles, with some open field lines emerging from lower latitudes. Proton temperatures reach about 3 MK, while the electron reach 1.5 MK, consistent with our previous analysis for the ideal dipole case. \par

\begin{figure} 
%\begin{center} 
%\epsscale{.70}
\plotone{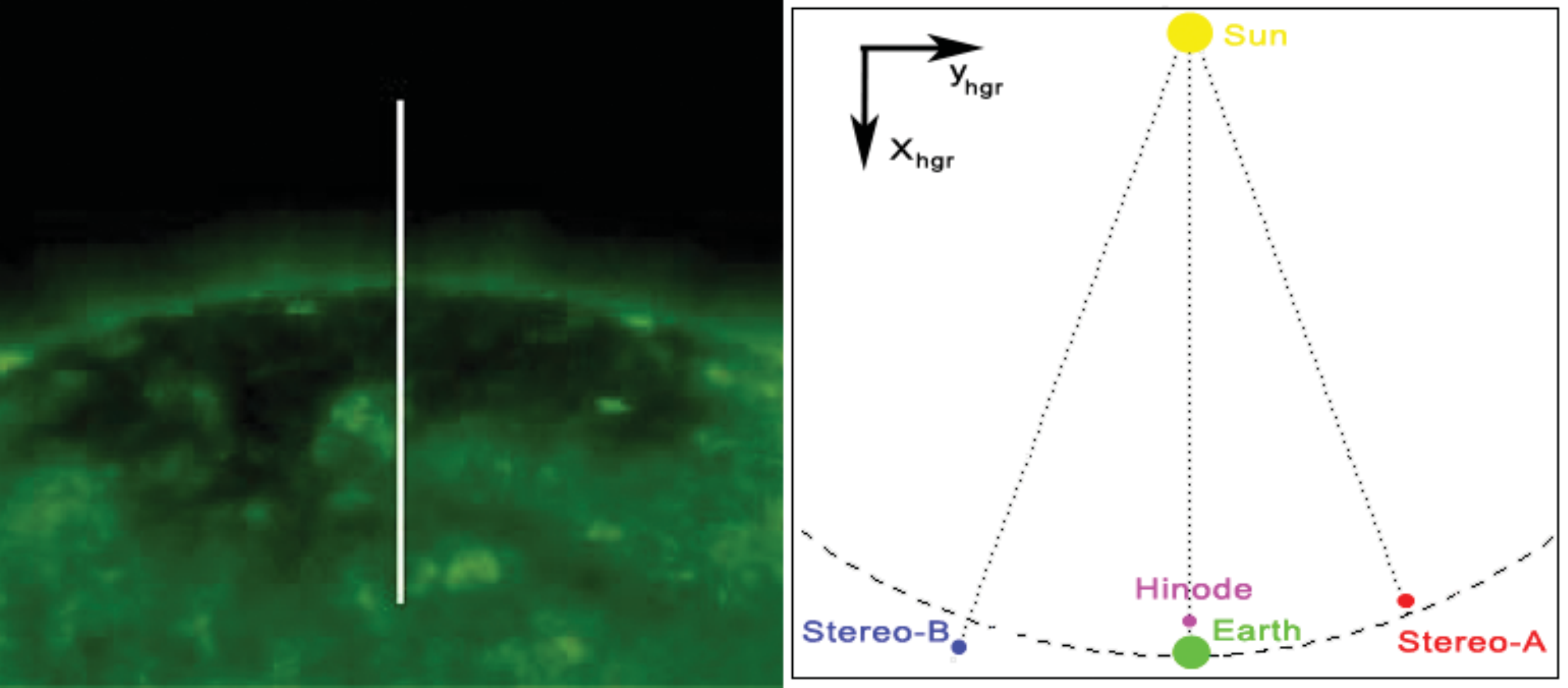} 
\caption{\small \sl Left panel: Location of the EIS slit used to observe coronal hole spectra for electron temperature and density diagnostics. The slit is overlaid on an EUV image from the Extreme ultraviolet Imaging Telescope (EIT) on board SOHO, taken on November 16, 2007. Right panel: positions of the STEREO-A, STEREO-B and Hinode (Solar-B) spacecraft for November 17, 2007, projected on the x=0 plane of the Heliographic Inertial (HGI) coordintate system.\label{F:eis_slit_sat_pos}} 
%\end{center} 
\end{figure}

\begin{figure} 
%\begin{center} 
%\epsscale{.70}
\plotone{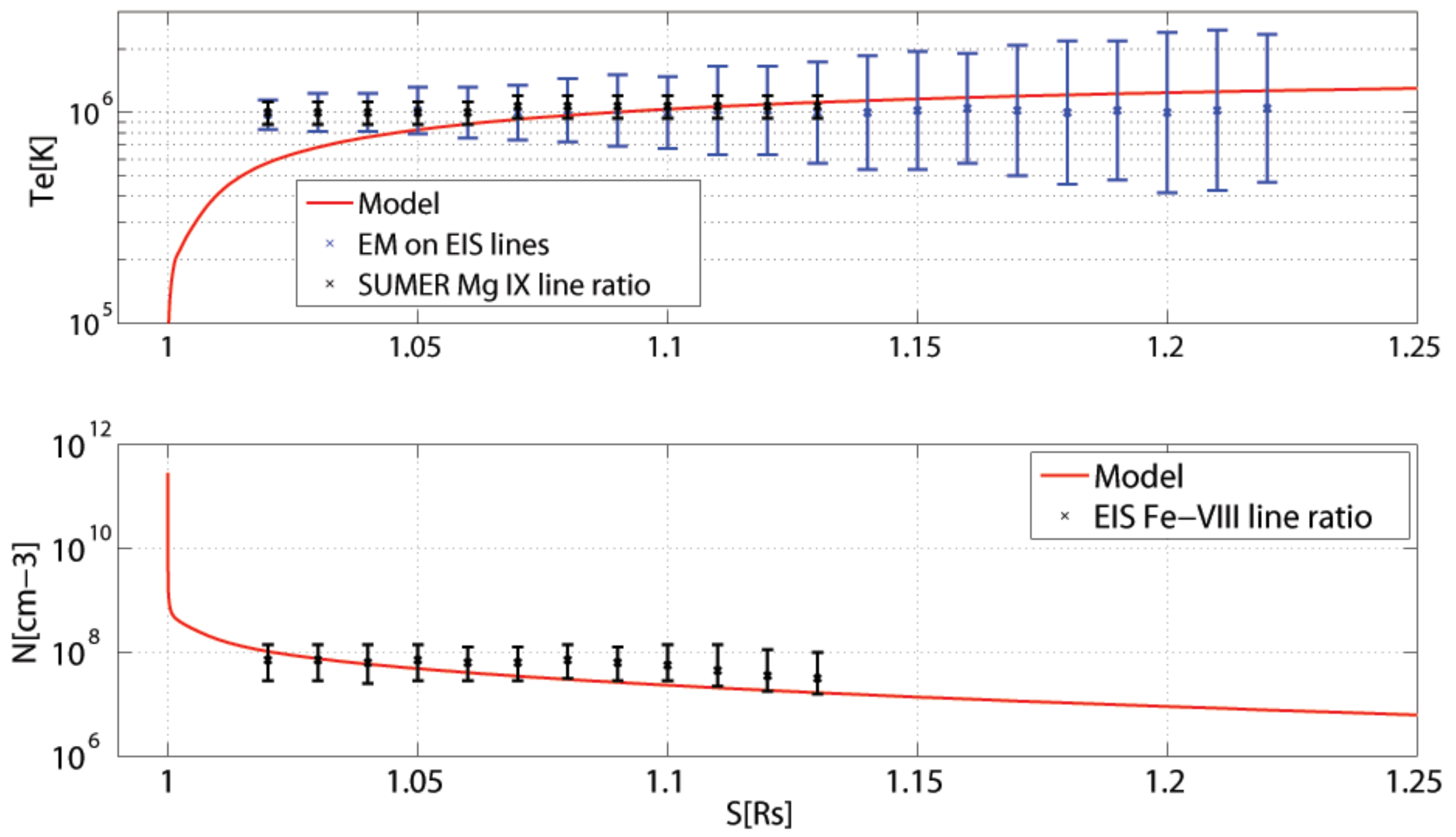} 
\caption{\small \sl Observed vs. predicted electron temperature (top panel) and density (bottom panel) radial profiles. The electron temperature was calculated using two methods  Mg IX line ratio (blue curve) measured by SUMER, and the EM loci method (black curve) on EIS spectral lines. The density was calculated from the Fe-VIII line intensity ratio.  \label{F:eit_temp_dens}} 
%\end{center} 
\end{figure}

In order to compare the predicted temperature and density profiles in the corona to observations, we use spectral line intensities measured by the EUV Imaging Spectrometer (EIS) on-board the Hinode (Solar-B) spacecraft \citep{culhane2007} and the Solar Ultraviolet Measurements of Emitted Radiation (SUMER) instrument on board SOHO \citep{wilhelm1995}. EIS observations were performed along the slit shown in Figure \ref{F:eis_slit_sat_pos} during November 16, 2007. The SUMER slit was placed at the same position as the EIS slit in the east-west directions, but stretched radially from $1.0$ to $1.3R_s$. The density was calculated using the EIS Fe-VIII line intensity ratio.  The electron temperature was calculated using two methods: Mg IX line intensity ratio from SUMER, and the EM loci method applied to EIS lines, as described in \citet{landi2008}. It should be noted that the spectral intensities used in this calculation are integrals along the line-of-sight. In order to recover the density and temperature profiles responsible for the emission it was assumed that the coronal hole plasma is optically thin in these wavelengths. Observational data below $1.02R_s$ was discarded due to the presence of spicule plasma, which is not optically thin. The model results were extracted along a magnetic field line passing through the center of the coronal hole and overlapping the slit. The profile was remapped in order to account for the artificial broadening of the transition region, as we described in Section (\ref{S:geometry}).  The transition region broadening affected results up to $1.02R_s$. Comparison of the observations to model results is shown in Figure \ref{F:eit_temp_dens}. The top panel shows the density, while the bottom panel shows the electron temperature. As can be seen, the model results agree rather well with the data above a distance of $1.05R_s$. The apparent disagreement between measured and predicted electron temperatures at altitudes lower than $1.05R_s$ is misleading. The line of sight to the location of the selected field line will in general pass through field lines that have their foot points at lower latitudes, and correspondingly the plasma flowing along these lines is at a higher altitude above the limb. Since the electron temperature increases with altitude, our temperature measurement is contaminated by hotter plasma contributing to the line of sight intensity.

\subsection{Multi-Point EUV and Soft X-Ray Images}
Full-disk emission images of the lower corona serve as an important diagnostic tool for global models. The photon flux in a given spectral line will in general depend on the electron density and temperature distribution along the line of sight to the detector, and therefor comparing model results to full-disk images in different spectral bands will allow us to test how well the predicted three-dimensional temperature and density distributions agree with the observations. In order to make the comparison, we must create synthetic line-of-sight images from the model results. In the most general case, this requires solving the full radiative transfer problem, which can be rather complex. For a first-approximation comparison, however, it is sufficient to assume the plasma is optically thin in the wavelengths under consideration. In this limit, the number of photons in a spectral band $i$, detected in a unit time at a given pixel in the imager, is given by:
\begin{equation}
\Phi_i = \int n_e^2 f_i(n_e,T_e)dl \quad [dN s^{-1}],
\end{equation}
where $dl$ is a path length along the line-of-sight, $n_e$ and $T_e$ are the electron density and temperature, respectively, and $f_i(n_e,T)$ is the instrument response function in that band. $\Phi_i$ is measured in units of number of photons per second, $dNs^{-1}$. Since our model does not simulate the wind-induced departures from ionization equilibrium, the response functions $f_i$ are constructed from the CHIANTI 7.1 atomic database \citep{dere1997,landi2013}, based on coronal elemental abundances \citep{feldman1992}, and assuming ionization equilibrium obtained from the ionization and recombination rates appearing in \citet{landi2013}.\par
We here compare our model results to both EUV and soft X-ray images. We use EUV images obtained by the Extreme Ultraviolet Imager (EUVI) on board the two STEREO spacecraft \citep{howard2008}. For soft X-ray images, we use the X-ray Telescope (XRT) on board the Hinode (Solar-B) mission \citep{kano2008,matsuzaki2007}. Both observed and synthesized images were taken around 2007-11-17, 01:00:00 UTC, which is approximately at the middle of the Carrington rotation, making the comparison to a steady-state solution most appropriate. At the time of observation, STEREO-A and STEREO-B were separated by about 40.5 in heliographic longitudes, with Hinode's position roughly in between them, along the Sun-Earth line. This set-up allows for a multi point-of-view model-data comparison. The respective locations of the observatories are shown in Figure \ref{F:eis_slit_sat_pos}.  In preparing the observed images from the raw data, including calibration, noise reduction and normalization of the photon flux by the exposure time, we used the SolarSoft (SSW) package written in IDL \citep{freeland1998}.\par

\begin{figure} 
%\begin{center} 
\epsscale{.90}
\plotone{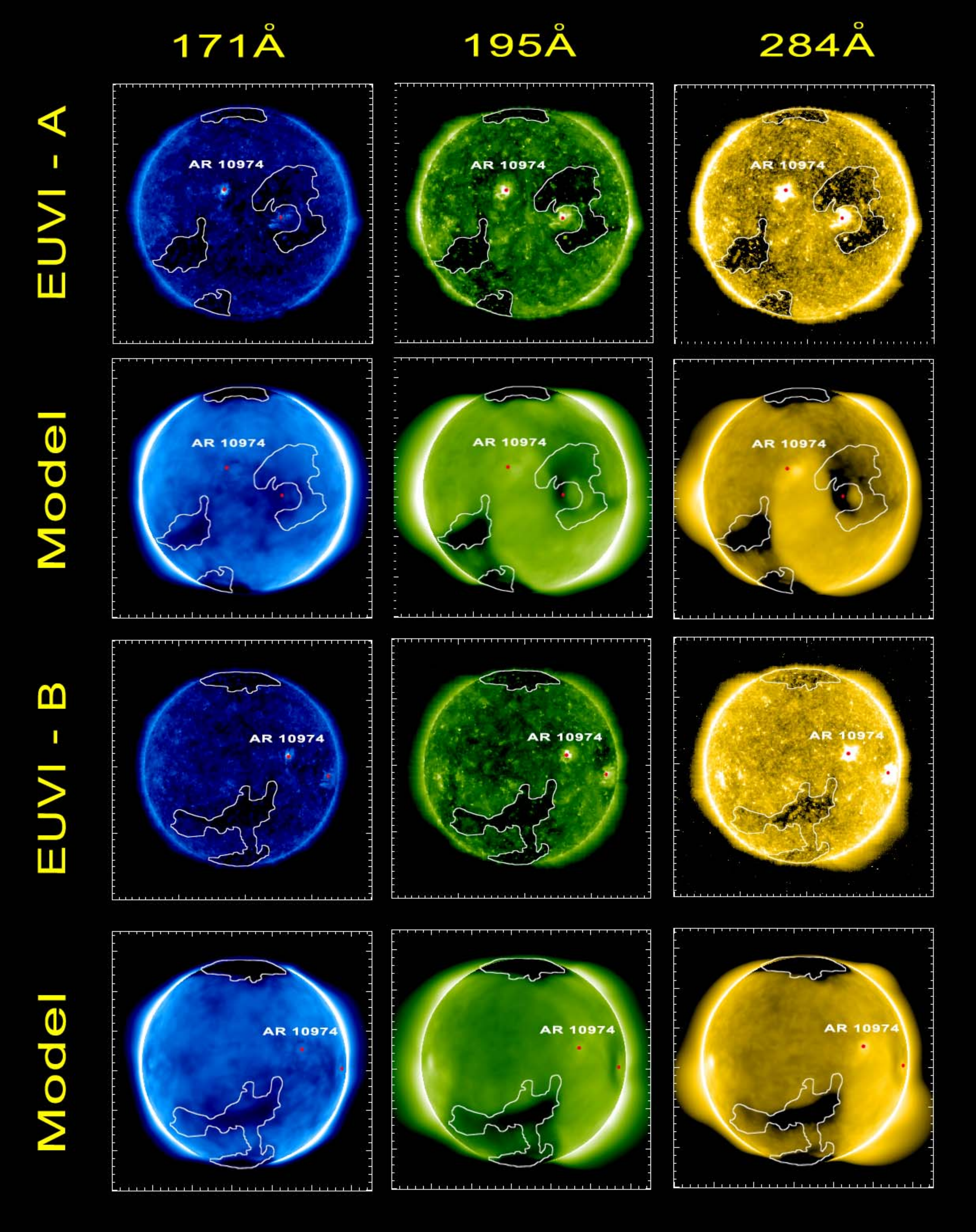} 
\caption{\small \sl STEREO/EUVI images vs. synthesized images in three different bands using the S1 filter. Top two panels: observations and synthesized images for EUVI-A (STEREO Ahead). Bottom two panels: observation and synthesized images for EUVI-B (STEREO Behind). The spacecraft location at the time of observation is shown in the right panel of figure (\ref{F:eis_slit_sat_pos}).\label{F:2063_euvi}} 
%\end{center} 
\end{figure}

\begin{figure} 
%\begin{center} 
\epsscale{0.8}
\plotone{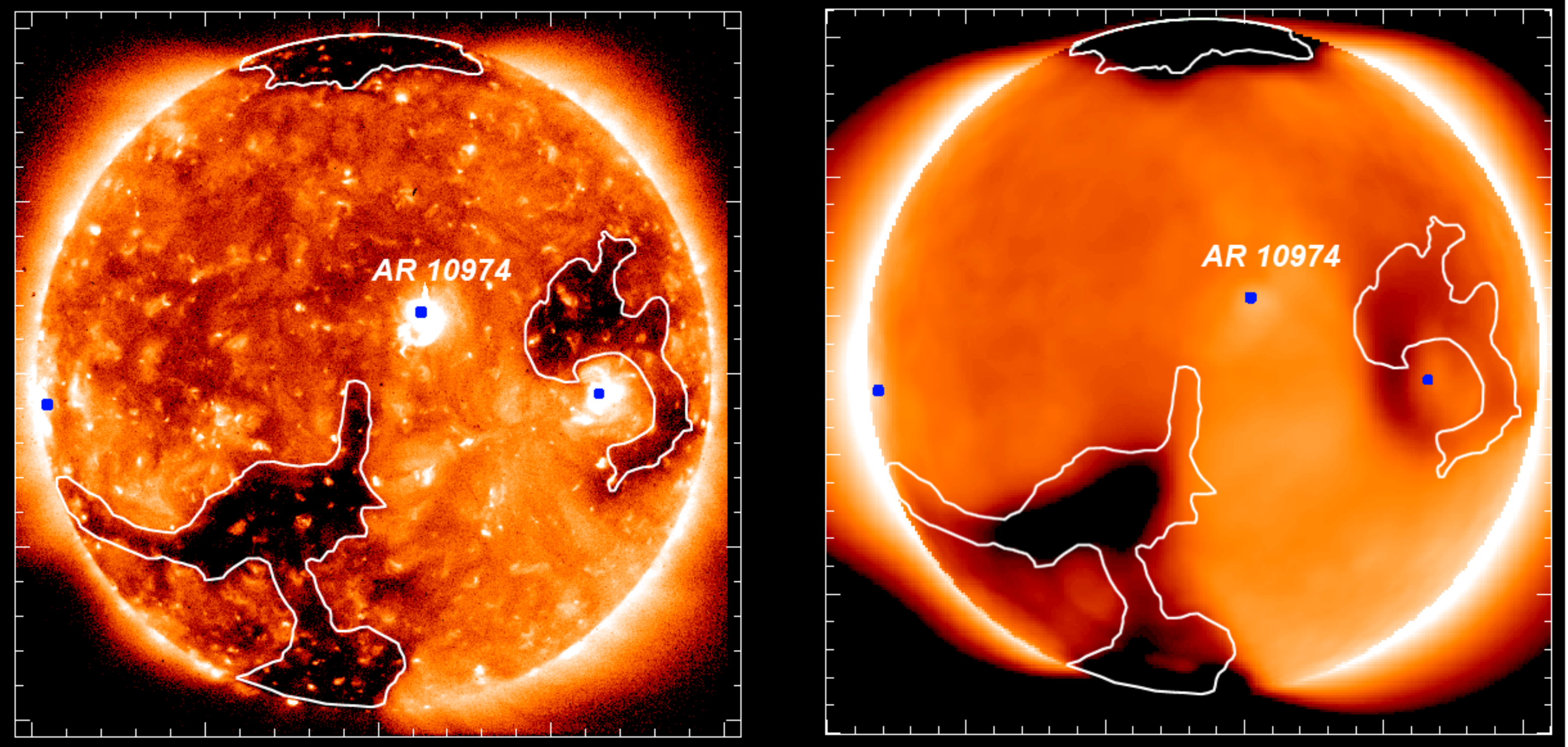} 
\caption{\small \sl Observed (left panel) and synthesized (right panel) images for the Hinode/XRT instrument, using the Al-Poly filter. The location of the Hinode spacecraft at the time of observation is shown in the right panel of figure (\ref{F:eis_slit_sat_pos}). \label{F:2063_xrt}} 
%\end{center} 
\end{figure} 

For EUVI-A and EUVI-B comparison, we use the $171${\AA}, $195${\AA} and $284${\AA} wavelengths, which are dominated by the ions Fe-IX, Fe-XII, and Fe-XV, respectively. The corresponding temperature ranges are $1MK$, $1.4MK$, and $2.2MK$. The images were obtained using the S1 filter, and the response tables for the synthesized images were calculated accordingly. The comparison is shown in Figure \ref{F:2063_euvi}. Each column corresponds to a different spectral band, with temperature increasing from left to right. The top two rows show observed and predicted emission for STEREO-A, while the two bottom rows show a comparison for STEREO-B. Figure \ref{F:2063_xrt} shows the comparison of model results to the XRT soft X-ray image, taken using the thin Al-poly filter, which is most sensitive to temperatures between $2MK$ and $10MK$ \citep{golub2007}. The left panel shows the observed image, while the right panel shows the synthesized image.\par
We marked the location of the active region and other bright features on the solar disk in both observed and synthesized images. Note that there is only a single active region with a NOAA designation for that time period. Although some traces of the active region appear in all synthesized images, the model best captures the intensity of this region in the $284${\AA} band. In all bands, the active region is fainter compared with the observations. This suggests that in the modeled active regions, the material possessing the corresponding temperatures is not dense enough to produce sufficient radiative power. This discrepancy between observed and modeled active regions can be attributed to the fixed boundary conditions used in this simulation. First, the magnetic field at the inner boundary was derived from a synoptic magnetic map, constructed from disk-center observations acquired over an entire Carrington rotation. Such a map might not reflect the instantaneous magnetic field strength that exists at the moment of the observations, especially in the highly variable active regions. In addition, in the real corona and chromosphere, the high heating rates in the active region will lead to heat being conducted down to the chromosphere, resulting in chromosphere evaporation, which will cause more plasma to flow up into the active region loops (c.f. \citet{klimchuk2006}). Such a process is completely absent from our model, since we have a fixed density at the inner boundary. A dynamic boundary condition should be considered if one wants to more realistically simulate active regions in a global model.\par 
 In order to see how well the model reproduces the overall topology, we manually trace the coronal hole boundaries on the observed images, and overlay the resulting contour on the synthesized images. As can be seen, the model correctly reproduced the location and approximate shape of the coronal holes. Although the overall topology agrees quite well, there are some discrepancies between the predicted coronal hole boundaries and the observed one. It is important to note all EUV imagers suffer from some degree of stray light scattering into the imaging plane. The stray light contribution to the detected intensity is negligible in the brighter regions of the image, but can contribute significantly in the fainter regions. \citet{shearer2012} found that stray light contamination in EUVI can reach up to $70\%$ for the EUVI instrument, resulting in observed coronal holes that are likely brighter than in reality. The topology is best captured by the soft X-ray case, which reveals the hotter, and therefore higher, layers of the corona. This trend suggests that the model better predicts the temperature structure at higher altitudes. \par 
Finally, we note that some of the smaller scale details are not captured by the model, which can be attributed to the following:  
1. Magnetogram accuracy and resolution: since the  magnetogram field is the only external input to the model, information that is not well captured in the magnetogram will not be passed to the model. 2. A steady-state solution with fixed magnetic field boundary conditions cannot capture transient phenomena. 
3. The MHD model cannot resolve small-scale physical processes.

\subsection{Solar Wind Structure up to 2AU and Comparison to In-situ Measurements} 
\begin{figure}
%\begin{center} 
\epsscale{.70}
\plotone{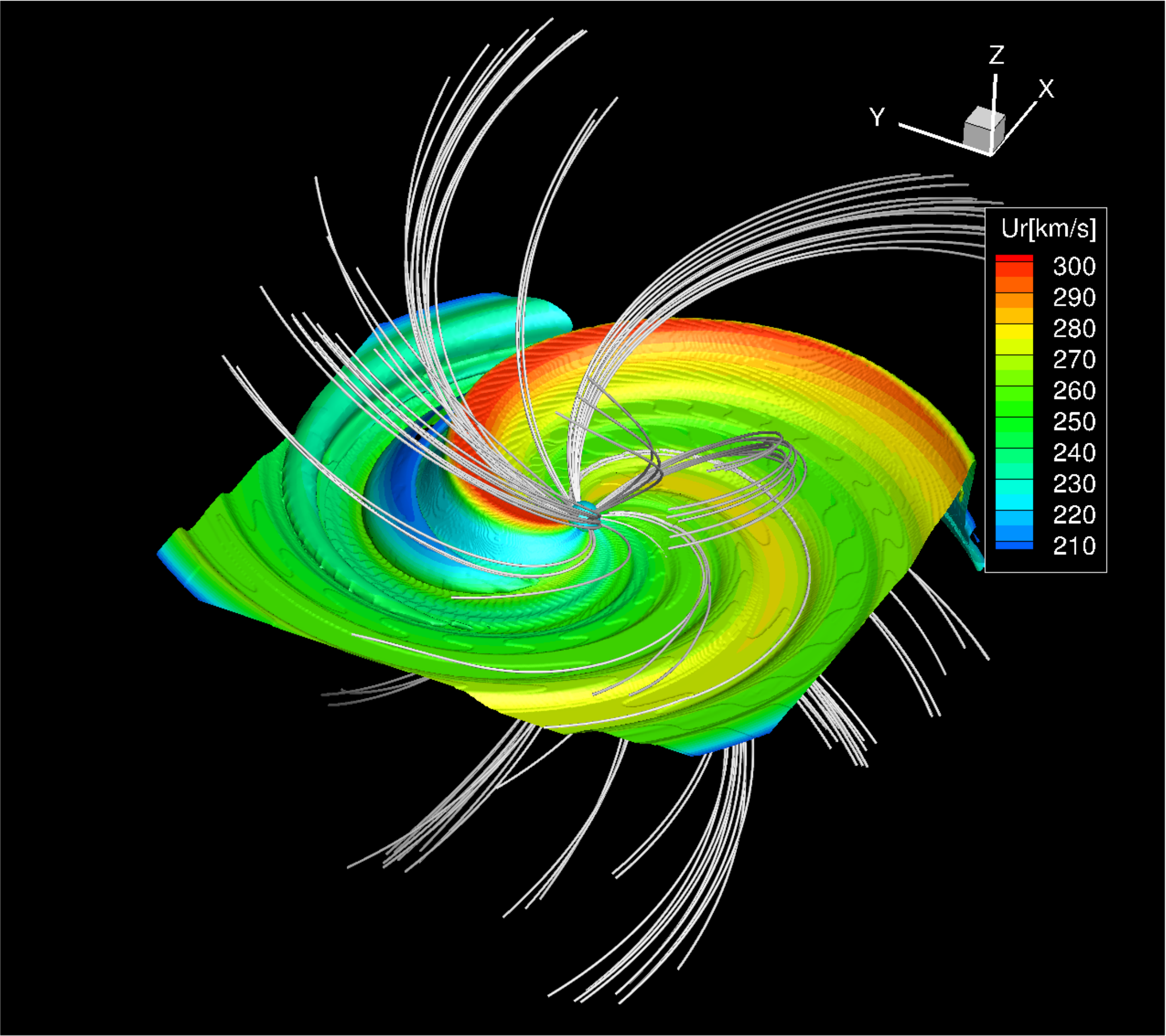} 
\caption{\small \sl Results for CR2063 up to a heliocentric distance of 2AU. Surface shows the location of the current sheet (where $B_r=0$), colored by the radial speed. Stream lines show the magnetic field. \label{F:2063_Ur3d}} 
%\end{center} 
\end{figure}

By coupling the solution in the SC component discussed in the previous sections to the IH component, we obtained a steady-state solution for CR2063 up to 2AU. Figure \ref{F:2063_Ur3d} shows the 3D structure of the solution, with magnetic field lines and the current sheet surface (where $B_r=0$) colored by the radial speed. The presence of interaction regions between the fast and slow streams is apparent.\par
\begin{figure}
%\begin{center} 
\epsscale{0.8}
\plotone{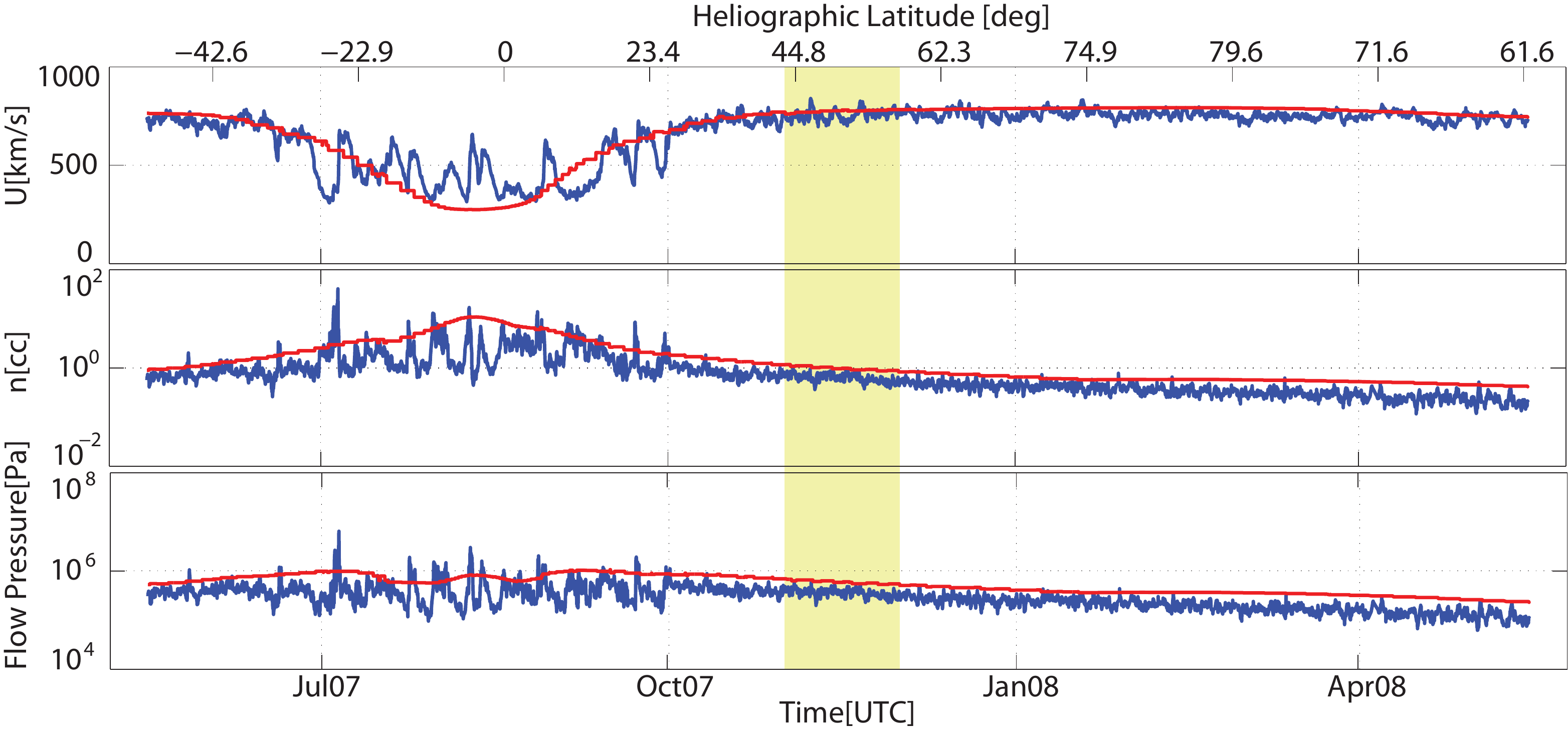} 
\caption{\small \sl Model-Data Comparison for CR2063 along Ulysses's orbit. Blue curves show Ulysses data and red curves show model data extracted along Ulysses's orbit. The shaded region denotes the period covered by the input magnetogram which was used to obtain the steady-state solution. The top panel shows the solar wind radial speed. The middle panel shows the proton density, while the bottom panel shows the proton dynamic pressure. \label{F:2063_Uly}} 
%\end{center} 
\end{figure}

One of the most important features of the solar wind is the latitudinal distribution of fast and slow solar wind streams, most comprehensively observed by the Ulysses spacecraft, orbiting the Sun in a nearly polar orbit. In order to examine how well the model reproduces these structures, we wish to compare our results to Ulysses measurements covering as wide a latitudinal range as possible. This requires an observation period much larger than a single Carrington Rotation, but since CR2063 took place within solar minimum, the latitudinal distribution of fast and slow wind streams does not change considerably from one Carrington Rotation to another. We therefore compare our model results to measurements taken from June 2007 to June 2008 (i.e. during a period of a year centered around the simulation time). Ulysses covered a latitude range between -55 to +80 degrees and heliocentric distances between 1.4 to 2 AU. Comparison of modeled wind speed, proton density, and dynamic pressure are shown in Figure \ref{F:2063_Uly}. The shaded region shows the period for which the magnetogram used as boundary condition was obtained. Note that this is a comparison between a steady state solution and a year worth of measurements, and therefore we do not expect to capture small scale or transient features. We also expect the agreement between the simulation and the observations to worsen as we move further away from the magnetogram time. What most concerns us here is to obtain the correct average properties of both the fast (high latitude) and slow (low latitude) wind. As can be seen from the top panel, the model has correctly captured the fast (~800 $km$ $s^{-1}$) and slow (~300 $km$ $s^{-1}$) wind speeds. The modeled proton density, shown in the middle panel, is only slightly higher than the observed one, and they are in very good agreement by order of magnitude. The bottom panel shows the wind dynamic pressure carried by the protons. At the heliocentric distances under consideration, this is the dominant energy component. As can be seen, here again the model and observations agree quite well.
\clearpage

\clearpage

 Thus, we have shown that our simulation has correctly predicted the distribution of wind acceleration in the inner heliosphere for CR2063, a solar minimum configuration. To complete this discussion, it would be instructive to examine the energy associated with the Alfv{\'e}n waves at these distances.  We refer back to the results shown in Figure \ref{F:polar_dU}, which has shown that the wave amplitude obtained for solar minimum (ideal dipole) case  is consistent with the results obtained by several observation campaigns.

\section{Conclusions}\label{S:discussion}
In this work, we presented and analyzed the AWSoM model, which is aimed at simulating the solar and heliospheric environment from the upper chromosphere to deep in the heliosphere within the extended-MHD approximation. In this model, a single heating mechanism is assumed: turbulent dissipation of Alfv{\'e}n waves. This mechanism is controlled by a simple set of three adjustable parameters, namely the chromospheric Poynting flux, the transverse correlation length, and a pseudo-reflection coefficient. \par 
Compared to previous global models, the wave dissipation mechanism assumed here is capable of treating both open and closed field line regions, and we do not need to a-priori determine whether a field line is open or closed. Rather, the open and closed magnetic structures emerge automatically and self-consistently with the distribution of solar wind speeds and coronal heating rates. This eliminates the need for empirical boundary conditions or geometric heating functions.\par 
We analyzed our choice of wave dissipation and adjustable parameters by simulating a steady-state solution for and ideal dipole configuration. We demonstrated that the sharp gradients in temperature and density between the chromosphere and the corona are reproduced, as well as the thermal differences between the polar coronal holes and the streamer belt.
As a further validation, we compared the predicted radial profile of wave energy to a large number of observations, ranging from the solar surface and up to 1AU. We found the predicted and observed profiles to be in good agreement.\par
Model-data comparison for CR2063 shows that the model simultaneously predicts the thermal structure near the Sun, as well as the flow properties of the solar wind at distances of 1-2 AU. This capability is a major step forward in global modeling of the entire chromosphere-to-wind system. We demonstrated this by comparing: 1. modeled electron density and temperature profiles to EIS and SUMER measurements 2. synthesized EUV and X-ray full disk images to observed ones, and 3. predicted solar wind properties to in-situ measurements obtained by Ulysses. \par
The two-temperature / extended MHD description better describes the energetics of the system compared to a single-temperature description. For the latter case, a higher Poynting flux would be required in order to sufficiently accelerate the fast wind to observed values. In the two-temperature case, the combined action of electron heat conduction and electron-proton thermal decoupling will modify the spatial distribution of heating and acceleration rates. The two-temperature description has the advantage of allowing us to extend model-data comparisons to a wider set of observables. In the present work, we tested the predicted electron properties against remote observations of the lower corona, and found them to be in good agreement at altitudes above 1.05 $R_s$. Predicted proton properties were compared to in-situ measurements in the solar wind. These were found to agree reasonably well, although a more complete thermodynamic description, such as the inclusion of collisionless heat conduction, might improve the results.\par
A robust model of the ambient solar corona and solar wind is a crucial building block in space weather prediction. The AWSoM model can be used to simulate eruptive events such as CMEs \citep{jin2013}, as well as to predict the location and properties of co-rotational interaction regions (CIR's) in the inner heliosphere. The small set of adjustable parameters can also provide a testing ground for various coronal heating models based on turbulent dissipation.\par
Finally, we mention possible ways to improve the present model. First, our model does not directly simulate wave reflections, and we assume a uniform reflection coefficient throughout the system. A more detailed and physics-based description of the wave dynamics is required to self-consistently determine the reflection coefficient from the local state of the plasma. Such a treatment will be included in a future publication (van der Holst et al. (2013, in preparation)). Second, the extended MHD description cannot account for the supra-thermal electron population. These electrons can carry a significant fraction of the thermal energy of the plasma, and affect the dynamics through the action of collisionless heat conduction (which becomes important at distances above 10 $R_s$). We plan to address these effects in forthcoming publications.

%% The \notetoeditor{TEXT} command allows the author to communicate
%% information to the copy editor.  This information will appear as a
%% footnote on the printed copy for the manuscript style file.  Nothing will
%% appear on the printed copy if the preprint or
%% preprint2 style files are used.

\acknowledgments
\begin{center}
\bf{Acknowledgments}\normalfont
\end{center}
The work presented in this paper was supported by the NSF CDI grant AGS-1027192 and NSF Space Weather Research grant AGS-1322543. We would like to extend our gratitude to Cooper Downs for useful suggestions and discussions. The simulations performed in this work were made possible thanks to the NASA Advanced Supercomputing Devision, which granted us access to the Pleiades Supercomputing cluster. We would like to thank NASA's CDAweb for supplying us with Ulysses observations.\par 
Analysis of radiative processes was made possible through the use of the CHIANTI atomic database. CHIANTI is a collaborative project involving the following Universities: Cambridge (UK), George Mason and Michigan (USA).\par 
We would like to thank the Hinode (Solar-B) team for supplying us with XRT and EIS data. Hinode is a Japanese mission developed and launched by ISAS/JAXA, with NAOJ as domestic partner and NASA and STFC (UK) as international partners. It is operated by these agencies in co-operation with ESA and NSC (Norway).\par

%% The reference list follows the main body and any appendices.%% Use LaTeX's 
% thebibliography environment to mark up your reference list.

\end{document}